\PassOptionsToPackage{unicode}{hyperref}
\PassOptionsToPackage{hyphens}{url}
\documentclass[
  12pt,
  a4paper,
]{article}
\usepackage{amsmath,amssymb}
\usepackage{setspace}
\usepackage{iftex,float}
\ifPDFTeX
  \usepackage[T1]{fontenc}
  \usepackage[utf8]{inputenc}
  \usepackage{textcomp} 
\else 
  \usepackage{unicode-math}
  \defaultfontfeatures{Scale=MatchLowercase}
  \defaultfontfeatures[\rmfamily]{Ligatures=TeX,Scale=1}
\fi
\IfFileExists{upquote.sty}{\usepackage{upquote}}{}
\IfFileExists{microtype.sty}{
  \usepackage[]{microtype}
  \UseMicrotypeSet[protrusion]{basicmath} 
}{}
\makeatletter
\@ifundefined{KOMAClassName}{
  \IfFileExists{parskip.sty}{%
    \usepackage{parskip}
  }{
    \setlength{\parindent}{0pt}
    \setlength{\parskip}{6pt plus 2pt minus 1pt}}
}{
  \KOMAoptions{parskip=half}}
\makeatother
\usepackage{xcolor}
\IfFileExists{xurl.sty}{\usepackage{xurl}}{} 
\IfFileExists{bookmark.sty}{\usepackage{bookmark}}{\usepackage{hyperref}}
\hypersetup{
  pdftitle={Cooperative Dilemmas in Rational Debate},
  pdfauthor={Toby Handfield, Julián Garcia, Christian Hilbe, Shang Long Yeo},
  hidelinks,
  pdfcreator={LaTeX via pandoc}}
\usepackage[top=20mm,left=25mm,right=25mm,heightrounded]{geometry}
\usepackage{longtable,booktabs,array}
\usepackage{calc} 
\usepackage{etoolbox}
\makeatletter
\patchcmd\longtable{\par}{\if@noskipsec\mbox{}\fi\par}{}{}
\makeatother
\IfFileExists{footnotehyper.sty}{\usepackage{footnotehyper}}{\usepackage{footnote}}
\makesavenoteenv{longtable}
\usepackage{graphicx}
\makeatletter
\def\maxwidth{\ifdim\Gin@nat@width>\linewidth\linewidth\else\Gin@nat@width\fi}
\def\maxheight{\ifdim\Gin@nat@height>\textheight\textheight\else\Gin@nat@height\fi}
\makeatother
\setkeys{Gin}{width=\maxwidth,height=\maxheight,keepaspectratio}
\makeatletter
\def\fps@figure{htbp}
\makeatother
\providecommand{\tightlist}{%
  \setlength{\itemsep}{0pt}\setlength{\parskip}{0pt}}
\newlength{\cslhangindent}
\newlength{\csllabelwidth}
\newlength{\cslentryspacingunit} 
\newenvironment{CSLReferences}[2] 
 {
  \setlength{\parindent}{0pt}
  \ifodd #1
  \let\oldpar\par
  \def\par{\hangindent=\cslhangindent\oldpar}
  \fi
  \setlength{\parskip}{#2\cslentryspacingunit}
 }%
 {}
\usepackage{calc}

\newcommand{\CSLLeftMargin}[1]{\parbox[t]{\csllabelwidth}{#1}}
\newcommand{\CSLRightInline}[1]{\parbox[t]{\linewidth - \csllabelwidth}{#1}\break}

\usepackage{amsmath,amsthm,amssymb,setspace,enumitem,epsfig,titlesec,verbatim,array,eurosym,multirow}
\usepackage{nicefrac}
\usepackage{mathtools}
\usepackage[sort&compress,numbers]{natbib}
\usepackage[footnotesize,bf]{caption}
\usepackage{standalone}
\usepackage{tikz}
\usepackage{subcaption}
\usepackage{hyperref}
\usepackage{tabularx}
\usepackage{booktabs}
\usepackage{blkarray}
\usepackage[ruled,vlined]{algorithm2e}
\smallskip 

\usepackage{tikz}
\usetikzlibrary{arrows}
\usepackage{minitoc}
\usepackage{graphicx}
\usepackage{hyperref}
\usepackage{subcaption}
\usepackage{multirow}
\usepackage{multicol}
\usepackage{standalone}
\newcommand{\vect}[1]{{\boldsymbol #1}}

\newtheoremstyle{plainCl1}
{12pt}
{12pt}
{\it}
{}
{\bfseries}
{}
{\newline}
{}

\theoremstyle{plainCl1}

\newtheoremstyle{plainCl2}
{0pt}
{0pt}
{}
{}
{\bfseries}
{.}
{0.1cm}
{}

\theoremstyle{plainCl2}
\newtheorem{Example}{Example}

\ifLuaTeX
  \usepackage{selnolig}  
\fi

\title{Cooperative Dilemmas in Rational Debate}
\author{Toby Handfield, Julián Garcia, Christian Hilbe, Shang Long
Yeo\thanks{JG and TH: Monash University; CH: IT:U; SLY: National University of Singapore. Corresponding author: toby.handfield@monash.edu.}}
\date{8 April 2025}

\begin{document}
\maketitle

\setstretch{1.3}
\frenchspacing

\begin{abstract}
    As an epistemic activity, rational debate and discussion requires cooperation, yet involves a tension between collective and individual interests. While all participants benefit from collective outcomes like reaching consensus on true beliefs, individuals face personal costs when changing their minds. This creates an incentive for each debater to let others bear the cognitive burden of exploring alternative perspectives. We present a model to examine the strategic dynamics between debaters motivated by two competing goals: discovering truth and minimizing belief revisions. Our model demonstrates that this tension creates social dilemmas where strategies that are optimal for individuals systematically undermine the collective pursuit of truth. Paradoxically, our analysis reveals that increasing debaters' motivation to seek truth can sometimes produce equilibria with worse outcomes for collective truth discovery. These findings illuminate why rational debate can fail to achieve optimal epistemic outcomes, even when participants genuinely value truth.
\end{abstract}

\normalsize

\newpage

\hypertarget{background}{%
\section{Background}\label{background}}

Rational debate is widely celebrated as an epistemic practice through
which individuals and communities advance knowledge and refine beliefs.
While some debates are artificially designed to be purely competitive,
most discussions offer opportunities for intellectual cooperation across
domains ranging from scientific inquiry to policy development. Yet even
as participants work toward shared epistemic goals, they face a
persistent tension: the collective aim of discovering truth may conflict
with the individual preference for maintaining stable beliefs. This
tension creates a potential cooperative dilemma where individual
rational choices might undermine collective epistemic achievements.

Substantial evidence indicates that human reasoning evolved primarily as
a social rather than individual cognitive adaptation. The architecture
of human reasoning appears specifically designed for exchanging and
evaluating arguments in group settings (1, 2), with even the fundamental
structures of logic potentially emerging from dialectical practices
rather than solitary contemplation (3). Cross-cultural studies
demonstrate remarkably consistent patterns of argumentative sensitivity
across diverse societies (4), while developmental research shows that
children evaluate argument quality from a young age (5). These findings
align with philosophical perspectives on social epistemology, which
recognize the essential role of testimony and collaborative inquiry in
knowledge acquisition (6--9). Contemporary work in this area highlights
the strategic dimensions of collective knowledge production,
acknowledging that epistemic cooperation involves navigating both shared
and competing interests (10--12).

The strategic aspects of debate create the conditions for what have been
termed ``epistemic dilemmas''---situations where individually rational
epistemic choices lead to collectively suboptimal outcomes (13). When
engaged in rational discourse, participants typically balance at least
two competing motivations: the desire to hold true beliefs and the
desire to minimize cognitive disruption by maintaining stable beliefs.
This creates a situation analogous to other social dilemmas, where
individual incentives may pull against collective benefit. While
individuals benefit from collectively identifying truth, each
participant might prefer that others do more of the cognitive work of
exploring alternative perspectives. These dynamics may help explain
various phenomena, including confirmation bias, polarization in public
discourse, and the varying speeds at which different epistemic
communities converge on accurate beliefs. Understanding these dynamics
requires formal analysis of how different argumentation strategies
interact and how they relate to both individual and collective epistemic
goals.

This work addresses three key questions about rational debate. First, we
develop a framework that models different debate strategies and explores
what makes someone a cooperative discussion partner. For instance, a
cooperative debater might help others discover truth while minimizing
unnecessary disruption to their beliefs, while an uncooperative one
might refuse to engage, or might needlessly force others to repeatedly
revise their positions. Second, we examine whether rational debate can
create strategic dilemmas: situations where debaters' individually
optimal choices lead to outcomes that frustrate everyone's goals. We
approach this question both from the perspective of individual debaters,
who care about both finding truth and maintaining stable beliefs, and
from the perspective of a hypothetical social planner who cares only
about the group's overall accuracy. Finally, we study which debate
strategies occur in equilibrium, depending on the motivations of the
debaters, and examine whether groups always achieve more accurate
beliefs when individuals are more strongly motivated to seek truth.

\hypertarget{model-and-results}{%
\section{Model and results}\label{model-and-results}}

Drawing on a framework developed by Gregor Betz (14), we model a
rational debate using a pair of debaters who occupy positions in logical
space. A position in logical space is simply an assignment of truth
values to the propositions being debated. So for instance, in a
3-proposition debate, there are 8 complete positions that might be
occupied, corresponding to the $2^3$ ways of assigning truth values to
the three propositions.

At any point in the debate, some subset of the logically possible
positions remain \emph{tenable}: these are the positions that have not
been refuted by any earlier arguments. Debaters take turns introducing
new arguments, which have the effect of reducing the space of tenable
positions.

An argument makes positions untenable in the following way: consider two
propositions, $p$ and $q$. There are four logically possible
combinations of truth values for these propositions. The argument
``$p$, therefore $q$'' does not tell us anything about what is the
case if $p$ is false, so it leaves the not-$p$ positions untouched.
What the argument does say is that if $p$ is true, then $q$ is true
also. So it renders untenable all the positions at which $p$ is true
and $q$ is false.

We adopt classical logic rather than non-monotonic reasoning systems,
meaning that once an argument eliminates a position, that position
remains eliminated regardless of what other propositions might be
introduced. While real-world argumentation often involves defeasible
reasoning (where $p \to r$ but $p \: \& \: q \to \neg r$), our focus
on the dialectical dynamics of debate justifies this simplification.
Future work could explore more complex logical frameworks, but for our
present purposes, classical logic provides a cleaner basis for
understanding the strategic choices debaters face.

If a debater's present position is removed by a new argument, they move
to a new position. We assume debaters are reluctant to revise beliefs
any more than necessary, so they adopt a position that requires changing
opinion on as few propositions as possible. Using a spatial metaphor, we
can say that they prefer to move the smallest possible distance in
logical space.

\begin{figure}
\centering
\includegraphics{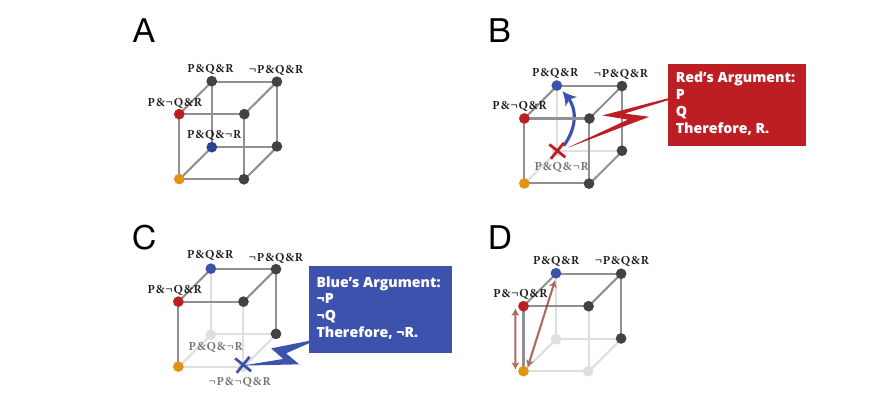}
\caption{Evolution of a debate over three propositions. The debater
positions are indicated by the red and blue nodes, while the truth is
the yellow node (A). Red introduces an argument which forces Blue to
adopt a new position (B). Blue produces an argument which affects an
unoccupied position (C). At the end of the debate, Red has 2 true
beliefs and Blue has one true belief (D).\label{deb-progress}}
\end{figure}

For purposes of visualization, it can be helpful to think of the space
of tenable positions as a graph, where each node is a tenable position,
which assigns a truth value to every proposition in the debate. We can
get a sense for the structure of the tenable positions by drawing edges
between any two positions that differ with respect to precisely one
proposition. In Figure \ref{deb-progress} we illustrate the evolution of
a brief debate over three propositions.

Additionally, we choose one node at random to be the factually true
position. In our baseline model, we initially assume that debaters only
produce valid arguments---those that don't eliminate the true position.
This represents ideally rational agents who respect logical constraints.
This approach establishes a baseline for understanding debate dynamics,
though we later extend the model to include an error parameter
($\alpha$) that allows for invalid arguments that might accidentally
rule out truth. It is also important to note that, unlike some
philosophical accounts of argumentation, our model treats propositions
as atomic and logically independent of each other. This deliberate
simplification allows us to focus on the strategic, dialectical aspects
of debate rather than on the substantive logical relationships between
propositions.

\hypertarget{the-process-of-debate}{%
\subsection{The process of debate}\label{the-process-of-debate}}

We model a debate with two agents, and $n$ propositions under debate.
The two debaters are initially assigned random positions.

At each timestep of the debate, we randomize whose turn it is to produce
an argument. If the debater's strategy can produce one or more valid
arguments, then it is further randomized which of those arguments will
be introduced. The space of tenable positions is then updated to reflect
the positions that have been eliminated by the argument. If either
debater's position is eliminated by an argument, that debater moves to a
remaining node that minimizes the number of belief changes required.

To evaluate how well different debate strategies help people discover
truth, we track what proportion of each debater's beliefs are actually
correct. For instance, if there are four propositions under debate, and
a debater correctly believes three of them but is wrong about the
fourth, their accuracy score would be 0.75. We are interested not just
in individual performance, but also in how well the debaters do as a
group. This collective accuracy -- which we calculate by averaging the
individual accuracy scores -- represents what a hypothetical impartial
observer would care about if their only concern was maximizing the
number of true beliefs across all participants. We also examine how
often debaters come to agree with each other, measuring consensus by
counting the proportion of issues on which they share the same view.
This helps us understand whether certain debate strategies are better at
promoting agreement, regardless of whether that agreement is on true or
false beliefs.

Real-world debates can end for many reasons: participants might run out
of time or energy, lose interest, or simply need to move on to other
activities. To reflect this in our model, we include a random chance
that the debate will end at any point. This chance is controlled by a
parameter for continuation probability ($\delta$) that we can adjust
to simulate shorter or longer debates. For example, with a low
$\delta$, debates tend to be brief, while a high $\delta$ allows for
extended discussions. Debates can also end in two other ways. First, if
the debaters have successfully eliminated all false positions through
their arguments, they will both arrive at the true position, and no
further progress is possible. Second, debates might reach a deadlock:
given the particular strategies the debaters are using, there might be
no more valid arguments available to either participant, even though
multiple positions remain tenable.

\hypertarget{argumentation-strategies}{%
\subsection{Argumentation strategies}\label{argumentation-strategies}}

Throughout a debate, participants follow consistent strategies for
constructing their arguments. Each strategy has two components: a rule
for choosing premises (the starting points of an argument) and a rule
for choosing conclusions (what the argument aims to establish). When
choosing premises, a debater has four options: they can use statements
that they themselves believe to be true, statements at least one of
which they believe to be false, statements their opponent believes to be
true, or statements at least one of which their opponent believes to be
false. Similarly, when choosing conclusions, they again have these same
four options. Combining these rules for premises and conclusions creates
sixteen possible strategies.

We label these choices using simple abbreviations: ``Self-Accept''
(\textbf{S+}) means using statements the debater believes,
``Self-Reject'' (\textbf{S--}) means using at least one statement the
debater disbelieves, ``Other-Accept'' (\textbf{O+}) means using
statements their opponent believes, and ``Other-Reject'' (\textbf{O--})
means using at least one statement their opponent disbelieves. For
example, a debater using the strategy ``\textbf{O+S--}'' builds
arguments by taking premises their opponent believes (\textbf{O+}) to
argue for conclusions that they themselves disbelieve (\textbf{S--}).

In addition to these 16 strategies, an additional strategy, ``Exit''
represents the complete refusal to enter debate, terminating the
exchange for both players. ``Exit'' captures a variety of
non-epistemically motivated ways to respond to the prospect of a debate.
(See Supplementary Information (\textbf{SI}) for full enumeration of strategies.)

To help understand the broad patterns in our model, we can categorize
the sixteen possible strategies into four groups: \emph{bold},
\emph{conservative}, \emph{compromising}, and \emph{refusenik}. While
this categorization is not essential to how the model works, it helps
illuminate the key differences in how combinations of strategies affect
debate outcomes. These four categories can be grouped into two higher
classes: monadic strategies (including bold and conservative) and dyadic
strategies (including compromising and refusenik).

\emph{Monadic} strategies are those where a debater builds arguments
using only one person's beliefs -- either their own beliefs or their
opponent's, but not both. Among these monadic strategies, we distinguish
two important types. The \emph{bold} strategies (\textbf{S+S-} and
\textbf{O+O-}) force at least one debater to change their position with
each argument, unless they already hold true beliefs. These strategies
are highly effective at finding truth but create significant disruption
by requiring frequent belief revision. In contrast, \emph{conservative}
strategies have the opposite property: at least one debater will never
be forced to change their position by a conservative strategy -- they
either use premises the debater rejects or argue for conclusions the
debater accepts. While these strategies minimize the uncomfortable
experience of changing one's mind, they do so at the cost of making it
harder to discover truth.

The remaining strategies combine beliefs from both debaters in
constructing arguments -- we call these ``\emph{dyadic}'' strategies
because they involve both participants' views. These strategies show
more varied patterns in their outcomes than the monadic strategies.
Within this group, two strategies stand out because they never force
anyone to change their mind: \textbf{S--O+} and \textbf{O--S+}. These
strategies work by taking premises that one debater disbelieves and
using them to argue for conclusions that the other debater accepts.
Since neither debater needs to revise any beliefs when faced with such
arguments, we call these the ``\emph{refusenik}'' strategies. We include
in this category the ``Exit'' strategy, where a debater simply refuses
to participate in the debate -- this achieves the same effect of
avoiding belief revision, but more directly, and denies the debate
partner the opportunity to make any arguments. All other dyadic
strategies can potentially cause at least one debater to change their
position, so we call them ``\emph{compromising}'' strategies. These
compromising strategies often lead to interesting dynamics: since they
incorporate both debaters' viewpoints, some of them help debaters
quickly reach agreement. However, this isn't universally true -- some
compromising strategies can actually keep debaters in disagreement for
long periods.

\begin{figure}
\centering
\includegraphics[width=0.95\textwidth,height=\textheight]{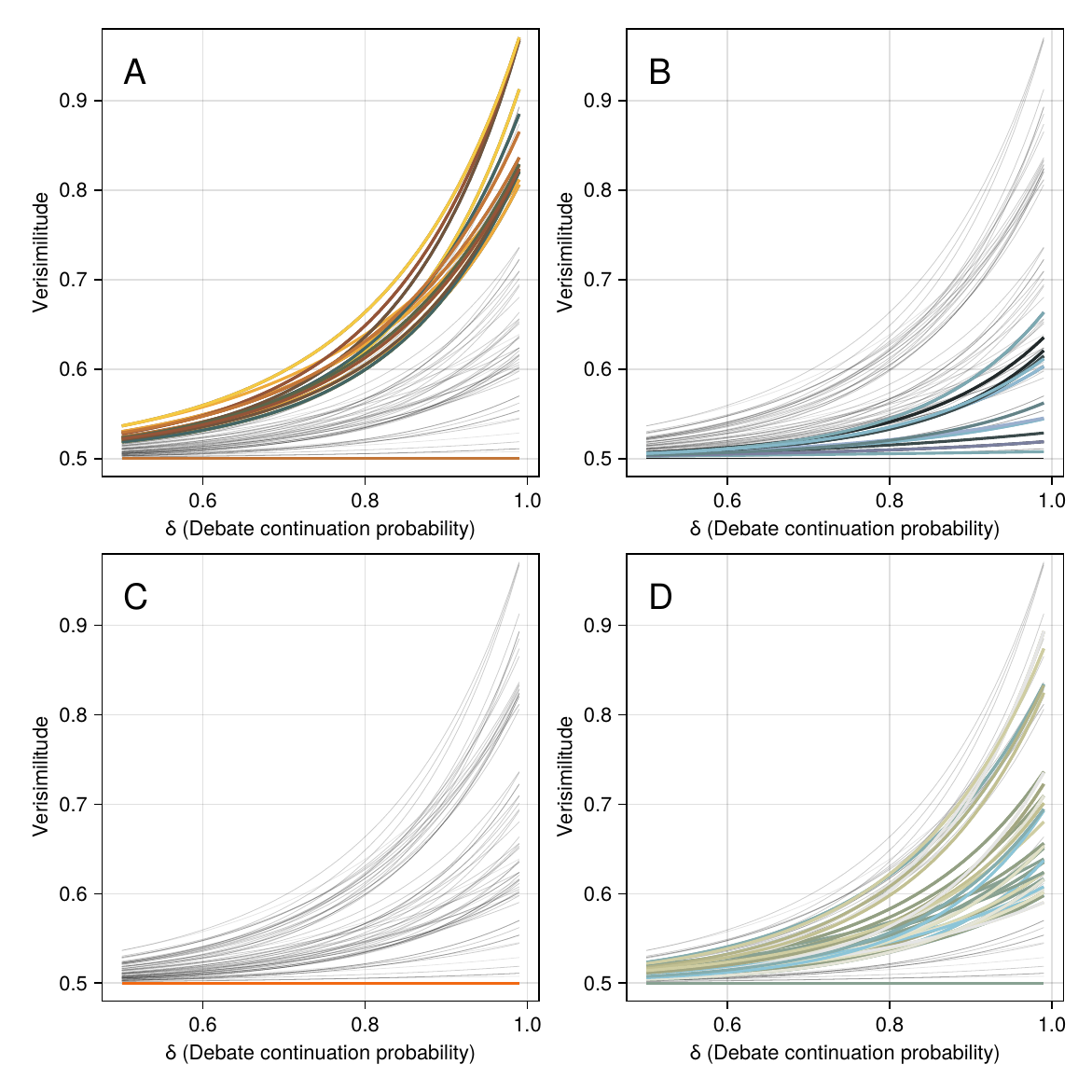}
\caption{Expected collective accuracy over debate length, for all
strategy profiles grouped by type. \textbf{A:} $\geq 1$ bold strategy
\textbf{B:} both conservative \textbf{C:} both refusenik \textbf{D:} $\geq 1$ compromising, none bold. $n = 3$. \label{all-verisim}}
\end{figure}

To illustrate how combinations of these strategies give rise to very
different collective epistemic outcomes, in Figure \ref{all-verisim} we
plot the expected levels of collective accuracy for various strategy
profiles (pairs of debater strategies). As the plots illustrate,
strategy profiles in which at least one strategy is bold tend to
maximize collective accuracy. When both strategies are conservative, or
both are refusenik, the collective epistemic performance is much worse.
The final category of profiles -- where at least one strategy is
compromising and no strategy is bold, is much more heterogeneous,
including both strong and weak outcomes for truth. (See \textbf{SI} for
analogous plots regarding the effect of various strategy profiles on
consensus.)

\hypertarget{debater-error}{%
\subsection{Debater error}\label{debater-error}}

To model the possibility that debaters sometimes are less than perfectly
rational, we extend our baseline model to include an error parameter
$\alpha$ which represents the probability that an invalid argument is
mistakenly (or indeed, deceptively) accepted. See \textbf{SI} for details on
precisely how the probability of argument selection is calculated when
$\alpha > 0$.

When invalid arguments are accepted, this entails that the truth is
removed. Once this happens the monotonic approach towards truth that is
observed in the baseline model is undermined: additional arguments can
take debaters further from truth (see Figure 3).

\begin{figure}
\centering
\includegraphics[width=0.95\textwidth,height=\textheight]{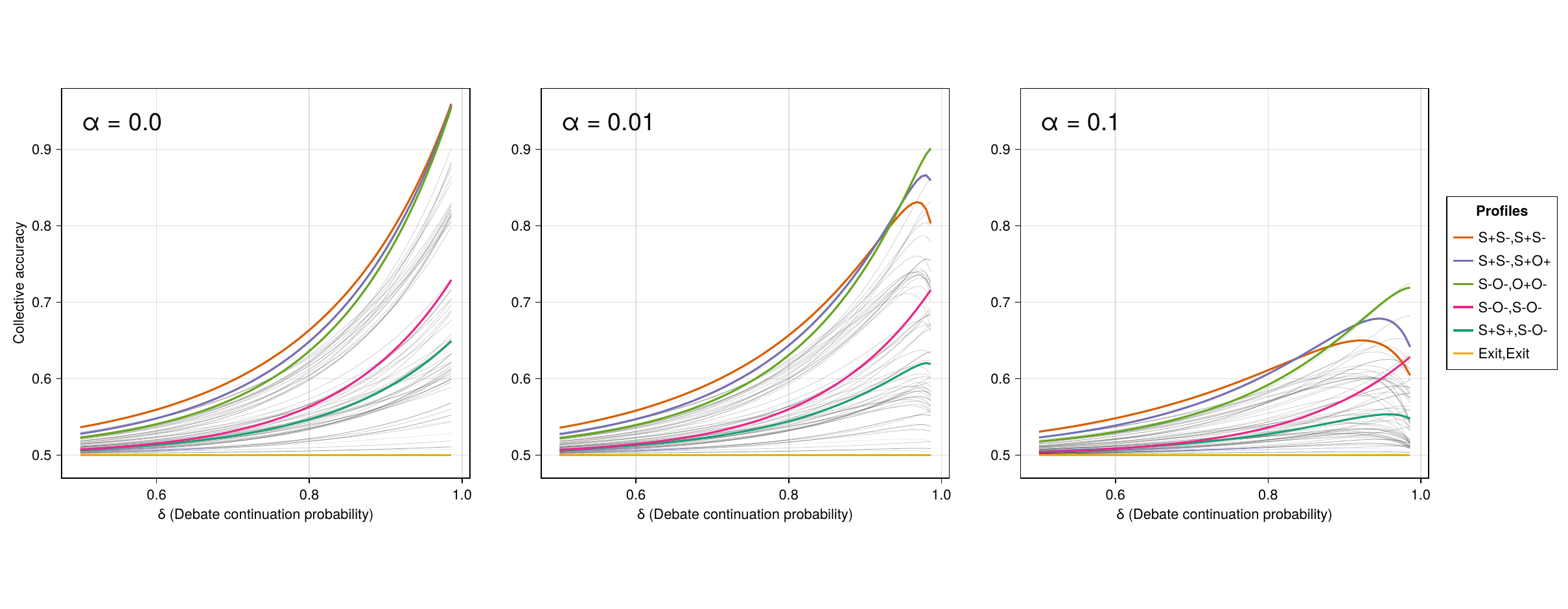}
\caption{Expected collective accuracy over debate length, for varying
levels of $\alpha$. Profiles from Table 1 are identified in the
legend. At $\alpha = 0$, the strategy profile
(\textbf{S+S--},\textbf{S+S--}) is truth maximizing for all levels of
$\delta$, but at non-zero $\alpha$, other, less argumentative
strategies, fare better in long debates.}
\end{figure}

\hypertarget{what-agents-want}{%
\subsection{What agents want}\label{what-agents-want}}

We assume agents want to have beliefs that are as close as possible to
the truth, but are reluctant to change their mind (accuracy, obstinacy).
These payoffs reflect a familiar tension akin to the ``explore/exploit''
tradeoff in learning systems: agents must weigh the potential benefits
of discovering new truths (exploration) against the comfort and
efficiency of maintaining their current beliefs (exploitation) (15--17).
While not a direct parallel---our model concerns refutation rather than
active experimentation---both situations involve an agent deciding
whether to incur immediate costs for potential future epistemic gains.
Just as an algorithm might stick with a known reward source rather than
sample a new option that could be better, our debaters may resist
changing their beliefs even when doing so might lead them closer to the
truth.

We make the tradeoff between these motivations precise by using the
following utility function:

\[u(v,d) = w\cdot v - d\]

$v \in [0,n]$ is the number of beliefs that the agent has which are
true by the end of the debate. $d$ is the aggregate number of times
that the debater changed their mind on any particular proposition. $w$
is a term to reflect the weight the agent places on the desirability of
gaining a new true belief versus the undesirability of changing one's
mind: an agent weakly prefers a new true belief if they can obtain it
with $w$ or fewer changes of mind.

\hypertarget{the-ideal-cooperative-debater}{%
\subsubsection{The Ideal Cooperative
Debater}\label{the-ideal-cooperative-debater}}

To begin getting an intuition for what a cooperative argumentation
strategy is, consider an ideally altruistic debate partner. Such a
partner would enhance the accuracy of one's beliefs while minimizing
unnecessary changes of mind. They would systematically eliminate false
positions from logical space, preserving only the participant's current
position and the objectively true position. Once this stage is reached,
a final argument could target the participant's false beliefs, allowing
for maximum truth acquisition with minimal cognitive disruption. This
scenario effectively allows the beneficiary to free-ride, as their
partner does the challenging work of eliminating false options, leaving
the free-rider to make only the most necessary belief changes to
maximize truth.

While our strategy space does not permit such finely targeted
dialectical altruism, this ideal serves as a useful limiting case.
Generally, agents in our model prefer partners who balance two traits:
they should minimize unnecessary refutations of one's position to avoid
excessive changes of mind, while also being epistemically efficient in
rapidly eliminating false positions. As we adjust model parameters to
represent debaters with greater concern for truth, the importance of
this second trait increases: non-disruptiveness becomes less crucial,
while epistemic efficiency gains prominence.

The strategic landscape also changes when debates are expected to be
short-lived. In brief exchanges, the ideal approach of systematically
ruling out all false positions before altering one's own beliefs becomes
less practical because the debate might end prematurely. As a result, a
debater who places a high value on discovering the truth might actually
prefer a partner who challenges their beliefs more aggressively. While
this approach may lead to more frequent changes of mind, it also offers
the potential for rapid improvements in the accuracy of one's beliefs.

\hypertarget{formal-definition-of-cooperation-and-cooperative-dilemmas}{%
\subsubsection{Formal Definition of Cooperation and cooperative
dilemmas}\label{formal-definition-of-cooperation-and-cooperative-dilemmas}}

We can formally define what constitutes cooperation in debate as
follows, extending the framework of Peña and Nöldeke (18). A pair of
actions or strategies $C$ (cooperation) and $D$ (defection)
represent a \emph{cooperate--defect pair} iff:

\begin{enumerate}
\def\labelenumi{\arabic{enumi}.}
\item
  Mutual cooperation is preferred to mutual defection:
  $E(C,C) > E(D,D)$
\item
  The payoff from cooperation creates positive externalities:

  \begin{itemize}
  \tightlist
  \item
    The payoff to a cooperator against a cooperator is at least as great
    as the payoff to a cooperator against a defector: $E(C,C) \geq
    E(C,D)$
  \item
    The payoff to a defector against a cooperator is at least as great
    as the payoff to a defector against a defector $E(D,C) \geq
    E(D,D)$
  \item
    And at least one of these two inequalities must be strict
  \end{itemize}
\end{enumerate}

We say a game is a \emph{cooperative dilemma} if it has two actions
$C$, $D$, such that:

\begin{itemize}
\tightlist
\item
  $C$ is part of a socially optimal profile (i.e.~maximizes average
  payoff), and
\item
  $D$ is played in an equilibrium profile that is suboptimal, and
\item
  $C,D$ are a cooperate--defect action pair.
\end{itemize}

We can then categorize these cooperative dilemmas as prisoner's
dilemmas, stag hunts, or snowdrift games, being the three types of
cooperative dilemma possible in a $2 \times 2$ game.

With this definition, we survey the space of games, and observe that
there are numerous cooperative dilemmas. In Table 1 we give a
representative illustration, holding $\delta$ fixed at 0.8, and
sampling a variety of truth weights. Observe that at very low truth
weights, the cooperative actions are \emph{Compromise} strategies, and
the defect action is a \emph{Conservative} strategy. At higher truth
weights, the cooperative actions are \emph{Bold} strategies, and the
defect action is a \emph{Compromise} strategy. This pattern, whereby the
cooperative action in a cooperative pair is the more epistemically
productive of the pair, is observed across a wide range of parameters.

\begin{longtable}[]{@{}
  >{\raggedright\arraybackslash}p{(\columnwidth - 6\tabcolsep) * \real{0.2353}}
  >{\raggedright\arraybackslash}p{(\columnwidth - 6\tabcolsep) * \real{0.2941}}
  >{\raggedright\arraybackslash}p{(\columnwidth - 6\tabcolsep) * \real{0.2353}}
  >{\raggedright\arraybackslash}p{(\columnwidth - 6\tabcolsep) * \real{0.2353}}@{}}
\caption{Representative cooperative dilemmas arising across parameter
space, ($\delta$ = 0.8). For simplicity we only present the pure
strategy equilibria in the table, as these suffice to demonstrate the
existence of the relevant ``Defect'' strategies.}\tabularnewline
\toprule
\begin{minipage}[b]{\linewidth}\raggedright
\textbf{Regime}
\end{minipage} & \begin{minipage}[b]{\linewidth}\raggedright
\textbf{Social optima}
\end{minipage} & \begin{minipage}[b]{\linewidth}\raggedright
\textbf{Pure strategy equilibria}
\end{minipage} & \begin{minipage}[b]{\linewidth}\raggedright
\textbf{Cooperative dilemmas}
\end{minipage} \\
\midrule
\endfirsthead
\toprule
\begin{minipage}[b]{\linewidth}\raggedright
\textbf{Regime}
\end{minipage} & \begin{minipage}[b]{\linewidth}\raggedright
\textbf{Social optima}
\end{minipage} & \begin{minipage}[b]{\linewidth}\raggedright
\textbf{Pure strategy equilibria}
\end{minipage} & \begin{minipage}[b]{\linewidth}\raggedright
\textbf{Cooperative dilemmas}
\end{minipage} \\
\midrule
\endhead
Very low truth weight (2.5) & (S--O--,S--O--) or (O--S--,O--S--) &
(Exit,Exit) and (Exit,S+S+) & \textbf{Prisoner's Dilemma:} S--O-- v
S+S+ \\
Low truth weight (3.5) & (S--O--,O+O--) or (O--S--,S+S--) & (Exit,Exit)
and (S+S--,S+O+) & \textbf{Stag Hunt:} S--O-- v S+O+;
\textbf{Snowdrift:} O+O--/S+S-- v S+O+ \\
Moderate truth weight (4.5) & (O+O--,O+S+) or (S+S--,S+O+) & (Exit,Exit)
and (S+S--,S+O+) & \textbf{Snowdrift:} O+O--/S+S-- v S+O+ \\
High truth weight (9.5) & (O+O--,O+O--) or (S+S--,S+S--) & (Exit,Exit)
and (S+S--,S+S--) & No dilemmas \\
\bottomrule
\end{longtable}

\hypertarget{measuring-conflict-and-harmony-between-the-debaters}{%
\subsection{Measuring conflict and harmony between the
debaters}\label{measuring-conflict-and-harmony-between-the-debaters}}

If debate is a zero sum activity, then hoping for cooperation to occur
is naive. On the other hand, if debate is a pure coordination game then
cooperation should be relatively trivial. To convey the scope for
possible cooperation across parameter space, in Figure 4 we plot the
rank correlation between the players payoffs as a measure of the degree
of harmony/conflict between them (19). Where payoffs are strictly
anti-correlated, the game is competitive and we expect no cooperation,
where the payoffs are strictly aligned, cooperation is straightforward.
As the figure illustrates, there are always possibilities for
cooperation: the index is always positive -- but cooperation is
apparently easier in longer debates, and in a narrow range of truth
weights that varies depending on the length of the debate.

\begin{figure}
\centering
\includegraphics[width=0.8\textwidth,height=\textheight]{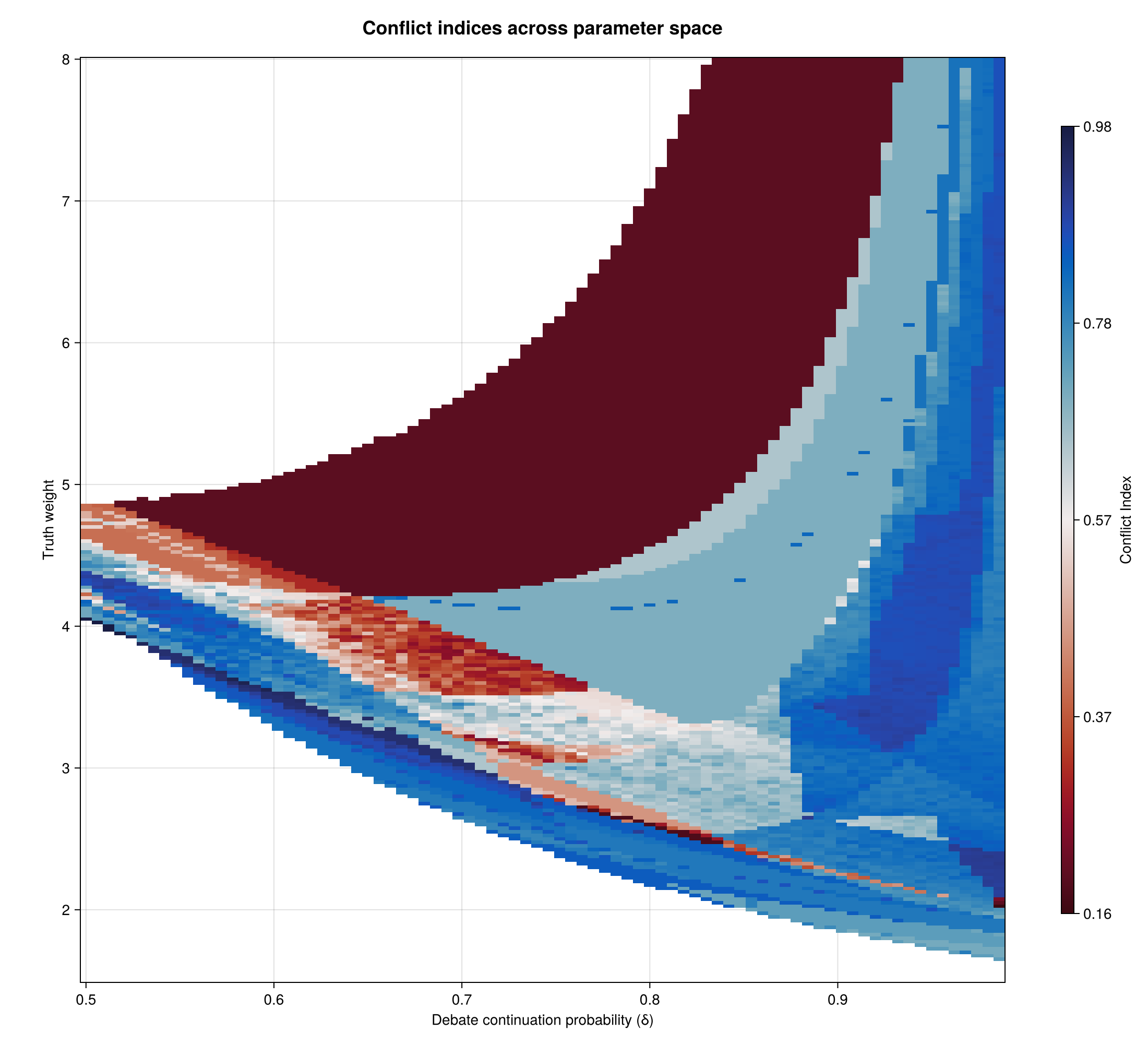}
\caption{Harmony/conflict index (rank correlation of players' payoffs)
across parameter space. To focus the analysis on the most relevant
payoff differences, we have first removed dominated strategies. In white
areas, the conflict index is undefined, because there is only one
dominant strategy.}
\end{figure}

\hypertarget{identifying-equilibria}{%
\subsection{Identifying equilibria}\label{identifying-equilibria}}

To study what behaviors are likely to emerge in actual debates, we use
the familiar Nash equilibrium concept from game theory. For small
debates ($n = 3$) we use numerical methods to calculate the expected
outcome of any two debating strategies being used, and assign utilities
to the outcome using the above utility function. For larger debates we
rely on simulations. We then identify pairs of strategies that are
mutual best responses: where neither debater has a reason to
unilaterally shift to an alternative strategy.

We prefer to use the Nash concept rather than the ESS concept (20),
which assumes symmetric equilibria, because we anticipate an important
possibility is that debaters might employ asymmetric strategy profiles,
where one uses a more cooperative strategy, while the other uses a less
cooperative one. Indeed, we observe this, particularly at intermediate
truth weights (see, e.g.~Table 1).

What equilibria we observe is highly sensitive to the parameters being
studied. Whether the debate is long or short, whether the debaters care
about truth, and whether the debaters make frequent errors, all change
the nature of the underlying game substantially.

There is often an abundance of equilibria at any given point in
parameter space. As a way of focusing our discussion on a salient set of
results, we introduce the idea of an \emph{epistemic planner} as an
equilibrium selection device. We assume that the epistemic planner
steers the debaters to the equilibrium that is most truth-conducive.
While this is an obvious idealization, it seems an apt choice to
highlight what emerges from the tension between ideally epistemic
concerns and the debaters' more mixed motives. As a robustness check we
also examine the results when the equilibrium selection uses a
utilitarian criterion, to simply maximize the debaters' payoffs (see
\textbf{SI}).

In Figure 5, upper panels, we plot the types of strategies that appear
in the truth-optimal equilibrium, at a range of parameters. In the lower
panels. We observe that at $\alpha = 0$, there are broadly four
regimes: an all-bold regime in the upper left; a refusenik regime in the
lower left; a bold/compromise regime that dominates the parameter space,
and a borderline region between the refusenik and the bold/compromise
regions, where conservative strategies appear.

Below we make some observations about these regimes and the way the
results change as we increase the error parameter $\alpha$.

\begin{figure}
\centering
\includegraphics{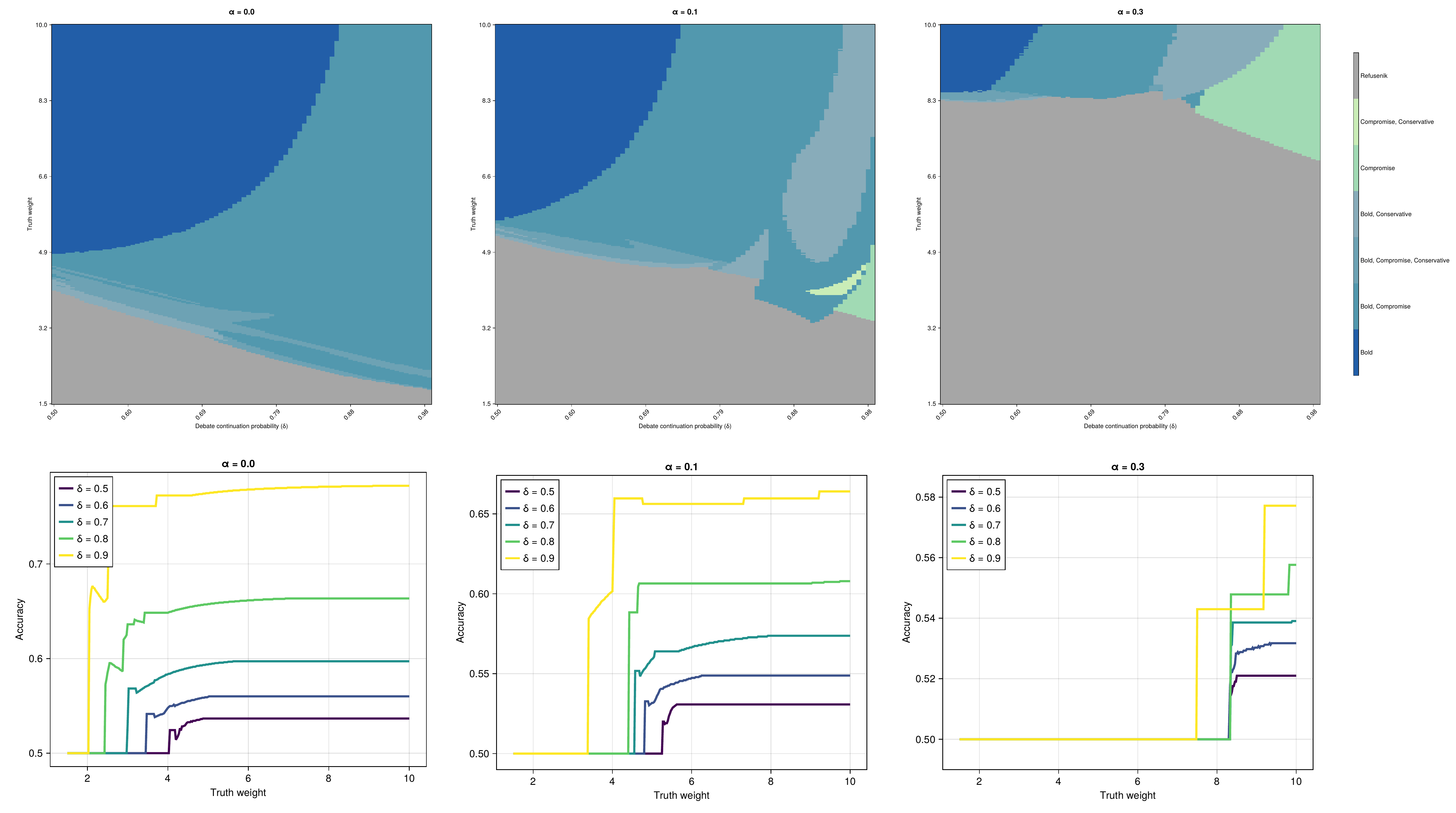}
\caption{\textbf{Upper panels:} overview of equilibrium types at various
levels of $\alpha$. \textbf{Lower panels:} Collective accuracy of
beliefs (normalized) as a function of debaters' truth-weight in the
utility function, at various levels of $\alpha, \delta$.}
\end{figure}

\hypertarget{agreeing-to-disagree-is-an-equilibrium-at-low-truth-weight-refusenik-regime}{%
\subsubsection{``Agreeing to disagree'' is an equilibrium at low truth
weight (Refusenik
regime)}\label{agreeing-to-disagree-is-an-equilibrium-at-low-truth-weight-refusenik-regime}}

In the upper left plot in Figure 5, where $\alpha = 0$, we observe
that there are broadly three regimes. Across all debate lengths, and
particularly when debates are short, if the truth-weight is low, the
equilibrium involves a refusenik strategy, especially Exit. Indeed,
manual inspection reveals that almost invariably these equilibria
involve (Exit,Exit). In such cases, the debaters refuse to produce
arguments, and so will have expected accuracy scores of 0.5. This is an
equilibrium, because at such low truth weights, debaters prefer to avoid
any change of mind, even if it means forgoing any epistemic gains. We
can describe this equilibrium as an ``agreement to disagree''.

\hypertarget{for-high-truth-motivated-agents-in-short-debates-there-is-a-unique-equilibrium-bold-regime}{%
\subsubsection{For high truth-motivated agents in short debates, there
is a unique equilibrium (Bold
regime)}\label{for-high-truth-motivated-agents-in-short-debates-there-is-a-unique-equilibrium-bold-regime}}

In the blue region indicating that the equilibrium involves only Bold
strategies, one strategy weakly dominates all others: \textbf{S+S--}.
The strategy profile in which both debaters use this strategy is the
only equilibrium, after removing dominated strategies. Even considering
all strategy profiles, it is also the truth optimal profile for almost
all values of delta, provided $\alpha=0$ (see Figure 3). This is thus
the most epistemically favourable region of parameter space.

\hypertarget{for-high-truth-motivated-agents-in-longer-debates-the-equilibrium-is-not-truth-optimal-boldcompromise-regime}{%
\subsubsection{For high truth-motivated agents in longer debates, the
equilibrium is not truth-optimal (Bold/compromise
regime)}\label{for-high-truth-motivated-agents-in-longer-debates-the-equilibrium-is-not-truth-optimal-boldcompromise-regime}}

At slightly lower truth motivation, or in somewhat longer debates,
however, the best response to a debater using \textbf{S+S--} is to adopt
a compromise strategy, such as \textbf{S+O+}. This sort of strategy is
capable of forcing the proponent to change their mind, but is not
guaranteed to do so: it will only force a change of mind if the
conclusion is a proposition on which the two debaters disagree.

This results in a modest loss of collective accuracy, and generates a
social dilemma that can be classed as a snowdrift game. The desire for
accurate beliefs is strong enough that if one debater is using a
compromise strategy, the other's best response is to use \textbf{S+S--}.
So we don't get mutual defection as an equilibrium. But while the two
debaters could solve the epistemic problem more efficiently if both
contributed maximally (both use \textbf{S+S--}), there is a temptation
for one agent to free ride on the labor of the other by adopting a
compromise strategy that is less likely to disrupt their own opinions.
Hence we get the asymmetric bold--compromise equilibrium.

\hypertarget{increasing-debater-error-increases-strategic-disengagement}{%
\subsubsection{Increasing debater error increases strategic
disengagement}\label{increasing-debater-error-increases-strategic-disengagement}}

As the second and third upper panels in Figure 5 dramatically
illustrate, if there are substantial error rates, refusenik strategies
come to dominate in equilibria. Indeed, closer inspection reveals that
Exit dominates all other strategies for large parts of parameter space.
This is straightforward to explain, as per the results illustrated in
Figure 3: because making arguments in these environments is much less
reliable as a way of approaching the truth, debaters need to be much
more motivated by truth-goals to make the risk of engaging in debate
worthwhile.

We also observe that in the regions where debaters do still engage, the
mixture of strategies changes: compromise and conservative strategies
arise much more often in equilibria, especially in longer debates. This
is also explicable in terms of the results shown in Figure 3, where
other strategy profiles overtake (S+S--,S+S--) in longer debates. Bold
strategies are distinctive in that they relentlessly produce new
arguments until there is only one position remaining. Other strategies
give rise to the possibility of stalemate, depending on the relative
positions of the debaters. In a long debate where error is possible,
this stalemate can be epistemically valuable: it is better to stop
producing arguments at some point, rather than to persevere and incur a
greater risk of removing the truth.

\hypertarget{collective-accuracy-does-not-increase-monotonically-with-accuracy-motives-of-the-debaters-borderline-region}{%
\subsubsection{Collective accuracy does not increase monotonically with
accuracy motives of the debaters (Borderline
region)}\label{collective-accuracy-does-not-increase-monotonically-with-accuracy-motives-of-the-debaters-borderline-region}}

In the lower panels of Figure 5, for $\alpha = 0.0, 0.1$, we observe
that collective accuracy does not monotonically increase with the
truth-weight in the debater's utility functions. This is paradoxical,
especially given that we are using an equilibrium selection criterion
that favours the truth when multiple equilibria are available.

This phenomenon corresponds to the borderline region of parameter space,
where Conservative strategies appear in the equilibria. In this region
of parameter space, we find no pure strategy equilibria where players
consistently use a single strategy. Instead, the game dynamics resemble
those of ``matching pennies'' or ``rock, paper, scissors,'' where no
single approach always wins.

In these situations, players in equilibrium use mixed strategies,
randomly choosing between different approaches. Counterintuitively, as
players become more invested in finding the truth (i.e., as the truth
weight increases), they use strategies that are less likely to help
their opponents reach accurate conclusions. This leads to less accurate
outcomes overall when the game reaches equilibrium. This is because of a
key requirement in mixed strategy equilibria: each player must adopt a
randomization strategy which make their opponent equally willing to use
any of their available strategies. As a player becomes more eager to
achieve a particular outcome (e.g.~true belief), their opponent must
therefore more frequently use strategies to frustrate that outcome. The
result is that the strategy a player most wants to use -- the one that
leads to their preferred outcome -- actually gets used less often as
their opponent cares more about that outcome.

\hypertarget{discussion}{%
\section{Discussion}\label{discussion}}

In other well-studied models of collective inquiry, knowledge advances
through both positive and negative evidence (21--23). Scientists might
discover a promising new experimental method, engineers might develop a
more efficient process, or researchers might find evidence supporting a
novel theory. These discoveries provide signals that help others infer
whether new approaches are better or worse than current practices. The
framework we study here, however, highlights a fundamental asymmetry in
rational argumentation: while we can conclusively prove that certain
combinations of beliefs are false, we can never definitively prove that
any particular theory is true through purely logical means.

This asymmetry echoes a profound insight from the philosophy of science,
most famously articulated by Karl Popper (24). While we can never
conclusively verify universal statements through observation -- no
matter how many white swans we observe, we cannot prove that all swans
are white -- we can definitively falsify such claims by finding a single
black swan. This limitation of inductive reasoning has far-reaching
implications for how knowledge progresses. Our model captures this
fundamental feature of rational inquiry: debaters can eliminate
positions from consideration by showing them to be logically
inconsistent, but they cannot definitively establish the truth of any
position through argument alone. While not all intellectual progress
follows this pattern -- scientific practice often relies on accumulating
positive evidence that provides strong but defeasible justification for
theories -- the logic of falsification plays a crucial role in both
formal argumentation and empirical investigation. Understanding how this
asymmetry affects strategic choices in debate may therefore shed light
on broader questions about the social dynamics of knowledge production,
from scientific collaboration to public discourse about complex policy
issues.

This asymmetry in rational argumentation also helps explain why agents
in our model exhibit inherent obstinacy. In settings where knowledge
advances through positive evidence, agents can be more confident that
changing their position represents genuine epistemic progress -- they
move toward better-justified beliefs based on accumulated evidence.
However, in a framework centered on deductive refutation, the situation
is markedly different. When forced to abandon a position by a valid
argument, agents have no way to determine which of the remaining
unrefuted positions is closest to truth. They might, in revising their
beliefs to accommodate a new argument, actually move further from the
truth. This fundamental uncertainty about the epistemic value of belief
revision makes reluctance to change one's mind -- what we call obstinacy
-- a rational response to the nature of deductive inquiry itself. This
echoes philosophical arguments (25) that when faced with evidence that
contradicts our beliefs, it is often rational to maintain most of our
belief system while making minimal adjustments to accommodate the new
information.

Our model offers a simplified way to study a fundamental tension in
collective truth-seeking: while society benefits when people work
together to discover truth, individuals often prefer to minimize changes
to their own beliefs. This connects to several important research
traditions. Evolutionary psychologists argue that human reasoning
evolved primarily as a social tool -- helping us to learn from others
while remaining vigilant to the threat of deception (2). Similarly,
sociologists of science have long studied how scientific communities
balance cooperation and competition in knowledge production (26). Our
model provides new insights into these dynamics. While our debaters
aren't trying to deceive each other or directly influence their
opponent's beliefs, they still face pressures that can lead them to
debate in ways that make finding truth harder for everyone. This happens
because their arguments shape which positions remain available to both
participants. This framework could be further generalized to help
explain several widespread phenomena: why people tend to favor evidence
supporting their existing beliefs (confirmation bias), why expert
communities sometimes maintain mistaken consensus (27), and why some
topics become increasingly polarized in public debate. The model might
also illuminate how social media and online forums affect the quality of
public discourse (28) and why scientific fields differ in how quickly
they converge on correct theories (29).

\hypertarget{analytic-methods}{%
\section{Analytic methods}\label{analytic-methods}}

\small

For small debates with $n = 3$ propositions, we represented the debate
process using a Markov chain. Each state in the Markov chain consists of
three elements: the set of tenable positions ($T$), and the current
positions of both debaters ($p'$ and $p''$). Since one of the
$2^n = 8$ positions must be the truth and always remains tenable,
there are $2^7$ possible sets of tenable positions. For each set, both
debaters must occupy tenable positions, yielding 2,816 possible states
overall.

We computed transition probabilities between states by considering: (\emph{i})
which debater makes the next argument (randomized), (\emph{ii}) what arguments
are feasible given their strategy, and (\emph{iii}) how these arguments affect
the tenable positions and potentially force position changes. For each
strategy profile, we calculated the expected payoffs by determining the
probability distribution over final states and the expected number of
belief revisions during the debate.

Debaters' payoffs follow $u(v,d) = w \cdot v - d$, where $v$ is the
accuracy (number of true beliefs at debate end), $d$ is the total
number of belief revisions, and $w$ represents the relative weight
placed on accuracy versus obstinacy. Debates end probabilistically with
parameter $\delta$ controlling expected debate length.

For larger $n$, where exact Markov analysis becomes computationally
infeasible (at $n = 4$, the state space grows to $\sim$2.5
million), we relied on simulations with ensembles of 200,000+ debates
per strategy profile.

We extended the baseline model by introducing an error parameter
$\alpha$ representing the probability that invalid arguments (those
eliminating the truth) are accepted. This parameter allowed us to
explore how debate outcomes change when perfect rationality is relaxed.
When we extend the model to allow invalid arguments, the transition
matrix must accommodate additional states in which the truth has been
removed from the set of tenable positions, leading to over 4,000 states.

For each combination of parameters ($\delta$, $w$, $\alpha$), we
identified Nash equilibria---strategy profiles where neither debater can
improve their payoff by unilaterally changing strategy. We then analyzed
how these equilibria and the resulting collective accuracy varied across
parameter space, with particular attention to cases where individual
incentives create cooperative dilemmas.

Equilibrium refinement: We iteratively removed dominated strategies to
simplify the strategic landscape. Because the Exit strategy induces ties
between all other strategies, no strategy is strictly dominated.
Although iterative removal of weakly dominated strategies is generally
avoided because it can violate order-invariance, in symmetric games this
is not the case. So we first iteratively removed weakly dominated
strategies, and if a multiplicity of equilibria remained we then chose
the equilibrium that was truth-optimal for closer study.

\hypertarget{references}{%
\section*{References}\label{references}}
\addcontentsline{toc}{section}{References}

\hypertarget{refs}{}
\begin{CSLReferences}{0}{0}
\leavevmode\vadjust pre{\hypertarget{ref-mercier_reasoning_2012}{}}%
\CSLLeftMargin{1. }
\CSLRightInline{H. Mercier, H. Landemore,
\href{http://onlinelibrary.wiley.com/doi/10.1111/j.1467-9221.2012.00873.x/full}{Reasoning
is for arguing: {Understanding} the successes and failures of
deliberation}. \emph{Political Psychology} \textbf{33}, 243--258
(2012).}

\leavevmode\vadjust pre{\hypertarget{ref-mercier_enigma_2017}{}}%
\CSLLeftMargin{2. }
\CSLRightInline{H. Mercier, D. Sperber, \emph{The {Enigma} of {Reason}}
(Harvard University Press, 2017).}

\leavevmode\vadjust pre{\hypertarget{ref-novaes_dialogical_2020}{}}%
\CSLLeftMargin{3. }
\CSLRightInline{C. D. Novaes, \emph{The dialogical roots of deduction:
{Historical}, cognitive, and philosophical perspectives on reasoning}
(Cambridge University Press, 2020).}

\leavevmode\vadjust pre{\hypertarget{ref-castelain_indigenous_2016}{}}%
\CSLLeftMargin{4. }
\CSLRightInline{T. Castelain, V. Girotto, F. Jamet, H. Mercier,
\href{https://doi.org/10.1016/j.evolhumbehav.2016.02.002}{Evidence for
benefits of argumentation in a {Mayan} indigenous population}.
\emph{Evolution and Human Behavior} \textbf{37}, 337--342 (2016).}

\leavevmode\vadjust pre{\hypertarget{ref-castelain_children_2018}{}}%
\CSLLeftMargin{5. }
\CSLRightInline{T. Castelain, S. Bernard, H. Mercier,
\href{https://doi.org/10.1111/infa.12202}{Evidence that
{Two}-{Year}-{Old} {Children} are {Sensitive} to {Information}
{Presented} in {Arguments}}. \emph{Infancy} \textbf{23}, 124--135
(2018).}

\leavevmode\vadjust pre{\hypertarget{ref-hardwig_epistemic_1985}{}}%
\CSLLeftMargin{6. }
\CSLRightInline{J. Hardwig,
\href{https://doi.org/10.2307/2026523}{Epistemic {Dependence}}.
\emph{The Journal of Philosophy} \textbf{82}, 335--349 (1985).}

\leavevmode\vadjust pre{\hypertarget{ref-goldman_knowledge_1999}{}}%
\CSLLeftMargin{7. }
\CSLRightInline{A. I. Goldman, \emph{Knowledge in a {Social} {World}}
(Oxford University Press, 1999).}

\leavevmode\vadjust pre{\hypertarget{ref-habermas1985theory}{}}%
\CSLLeftMargin{8. }
\CSLRightInline{J. Habermas, \emph{The theory of communicative action:
Volume 1: Reason and the rationalization of society} (Beacon press,
1985).}

\leavevmode\vadjust pre{\hypertarget{ref-longino1990-LONSAS-4}{}}%
\CSLLeftMargin{9. }
\CSLRightInline{H. E. Longino, \emph{Science as social knowledge: Values
and objectivity in scientific inquiry} (Princeton University Press,
1990).}

\leavevmode\vadjust pre{\hypertarget{ref-list_epistemic_2004}{}}%
\CSLLeftMargin{10. }
\CSLRightInline{C. List, P. Pettit,
{``\href{https://api.taylorfrancis.com/content/chapters/edit/download?identifierName=doi\&identifierValue=10.4324/9780203326886-8\&type=chapterpdf}{An
epistemic free-riding problem?}''} in \emph{Karl {Popper}: {Critical}
Appraisals}, P. Catton, G. MacDonald, Eds. (Routledge, 2004), pp.
128--158.}

\leavevmode\vadjust pre{\hypertarget{ref-dunn_epistemic_2018}{}}%
\CSLLeftMargin{11. }
\CSLRightInline{J. Dunn,
{``\href{https://doi.org/10.1093/oso/9780198779681.003.0014}{Epistemic
{Free} {Riding}}''} in \emph{Epistemic {Consequentialism}}, H. K.
Ahlstrom-Vij, J. Dunn, Eds. (Oxford University Press, 2018).}

\leavevmode\vadjust pre{\hypertarget{ref-zollman_theory_2020}{}}%
\CSLLeftMargin{12. }
\CSLRightInline{K. J. S. Zollman, The theory of games as a tool for the
social epistemologist. \emph{Philosophical Studies} (2020).}

\leavevmode\vadjust pre{\hypertarget{ref-mayo-wilson_independence_2011}{}}%
\CSLLeftMargin{13. }
\CSLRightInline{C. Mayo-Wilson, K. J. S. Zollman, D. Danks,
\href{https://doi.org/10.1086/661777}{The independence thesis: When
individual and social epistemology diverge}. \emph{Philosophy of
Science} \textbf{78}, 653--677 (2011).}

\leavevmode\vadjust pre{\hypertarget{ref-betz2013debate}{}}%
\CSLLeftMargin{14. }
\CSLRightInline{G. Betz, \emph{Debate dynamics: {How} controversy
improves our beliefs} (Springer Netherlands, 2013).}

\leavevmode\vadjust pre{\hypertarget{ref-holland1992adaptation_1992}{}}%
\CSLLeftMargin{15. }
\CSLRightInline{J. H. Holland, \emph{Adaptation in {Natural} and
{Artificial} {Systems}: {An} {Introductory} {Analysis} with
{Applications} to {Biology}, {Control}, and {Artificial} {Intelligence}}
(MIT Press, 1992).}

\leavevmode\vadjust pre{\hypertarget{ref-cohen_should_2007}{}}%
\CSLLeftMargin{16. }
\CSLRightInline{J. D. Cohen, S. M. McClure, A. J. Yu,
\href{https://doi.org/10.1098/rstb.2007.2098}{Should {I} stay or should
{I} go? {How} the human brain manages the trade-off between exploitation
and exploration}. \emph{Philosophical Transactions of the Royal Society
B: Biological Sciences} \textbf{362}, 933--942 (2007).}

\leavevmode\vadjust pre{\hypertarget{ref-hills_exploration_2015}{}}%
\CSLLeftMargin{17. }
\CSLRightInline{T. T. Hills, P. M. Todd, D. Lazer, A. D. Redish, I. D.
Couzin, \href{https://doi.org/10.1016/j.tics.2014.10.004}{Exploration
versus exploitation in space, mind, and society}. \emph{Trends in
Cognitive Sciences} \textbf{19}, 46--54 (2015).}

\leavevmode\vadjust pre{\hypertarget{ref-Pena2023}{}}%
\CSLLeftMargin{18. }
\CSLRightInline{J. Peña, G. Nöldeke,
\href{https://doi.org/10.1007/s13235-023-00524-9}{Cooperative dilemmas
with binary actions and multiple players}. \emph{Dynamic Games and
Applications} \textbf{13}, 1156--1193 (2023).}

\leavevmode\vadjust pre{\hypertarget{ref-tan_groups_2008}{}}%
\CSLLeftMargin{19. }
\CSLRightInline{J. H. W. Tan, D. J. Zizzo,
\href{https://doi.org/10.1016/j.socec.2006.12.023}{Groups, cooperation
and conflict in games}. \emph{The Journal of Socio-Economics}
\textbf{37}, 1--17 (2008).}

\leavevmode\vadjust pre{\hypertarget{ref-maynard_smith_game_1979}{}}%
\CSLLeftMargin{20. }
\CSLRightInline{J. Maynard Smith,
\href{https://doi.org/10.1098/rspb.1979.0080}{Game {Theory} and the
{Evolution} of {Behaviour}}. \emph{Proceedings of the Royal Society of
London B: Biological Sciences} \textbf{205}, 475--488 (1979).}

\leavevmode\vadjust pre{\hypertarget{ref-bala1998_learning}{}}%
\CSLLeftMargin{21. }
\CSLRightInline{V. Bala, S. Goyal,
\href{https://doi.org/10.1111/1467-937X.00059}{Learning from
{Neighbours}}. \emph{The Review of Economic Studies} \textbf{65},
595--621 (1998).}

\leavevmode\vadjust pre{\hypertarget{ref-Hong:2004ut}{}}%
\CSLLeftMargin{22. }
\CSLRightInline{L. Hong, S. E. Page, {Groups of diverse problem solvers
can outperform groups of high-ability problem solvers}.
\emph{Proceedings of the National Academy of Sciences of the USA}
\textbf{101}, 16385--16389 (2004).}

\leavevmode\vadjust pre{\hypertarget{ref-Zollman:2007comm}{}}%
\CSLLeftMargin{23. }
\CSLRightInline{K. J. S. Zollman,
\href{https://doi.org/10.1086/525605}{The {Communication} {Structure} of
{Epistemic} {Communities}}. \emph{Philosophy of Science} \textbf{74},
574--587 (2007).}

\leavevmode\vadjust pre{\hypertarget{ref-popper_lsd}{}}%
\CSLLeftMargin{24. }
\CSLRightInline{K. R. Popper, \emph{The logic of scientific discovery}
(Hutchinson, 1959).}

\leavevmode\vadjust pre{\hypertarget{ref-qui51}{}}%
\CSLLeftMargin{25. }
\CSLRightInline{W. V. O. Quine, Two dogmas of empiricism.
\emph{Philosophical Review} (1951).}

\leavevmode\vadjust pre{\hypertarget{ref-merton_sociology_1974}{}}%
\CSLLeftMargin{26. }
\CSLRightInline{R. K. Merton, \emph{The sociology of science:
Theoretical and empirical investigations} (University of Chicago Press,
1974).}

\leavevmode\vadjust pre{\hypertarget{ref-zollman_epistemic_2010}{}}%
\CSLLeftMargin{27. }
\CSLRightInline{K. J. S. Zollman,
\href{http://link.springer.com/article/10.1007/s10670-009-9194-6}{The
epistemic benefit of transient diversity}. \emph{Erkenntnis}
\textbf{72}, 17--35 (2010).}

\leavevmode\vadjust pre{\hypertarget{ref-bail_exposure_2018}{}}%
\CSLLeftMargin{28. }
\CSLRightInline{C. A. Bail, \emph{et al.}, Exposure to opposing views on
social media can increase political polarization. \emph{Proceedings of
the National Academy of Sciences} \textbf{115}, 9216--9221 (2018).}

\leavevmode\vadjust pre{\hypertarget{ref-shwed_temporal_2010}{}}%
\CSLLeftMargin{29. }
\CSLRightInline{U. Shwed, P. S. Bearman,
\href{https://doi.org/10.1177/0003122410388488}{The {Temporal}
{Structure} of {Scientific} {Consensus} {Formation}}. \emph{American
Sociological Review} \textbf{75}, 817--840 (2010).}

\end{CSLReferences}

\appendix
\newpage

\textbf{\huge Supplementary Information}

\newpage

\hypertarget{how-deductive-arguments-affect-logical-space}{%
\section{How deductive arguments affect logical
space}\label{how-deductive-arguments-affect-logical-space}}

\begin{figure}
\centering
\includegraphics{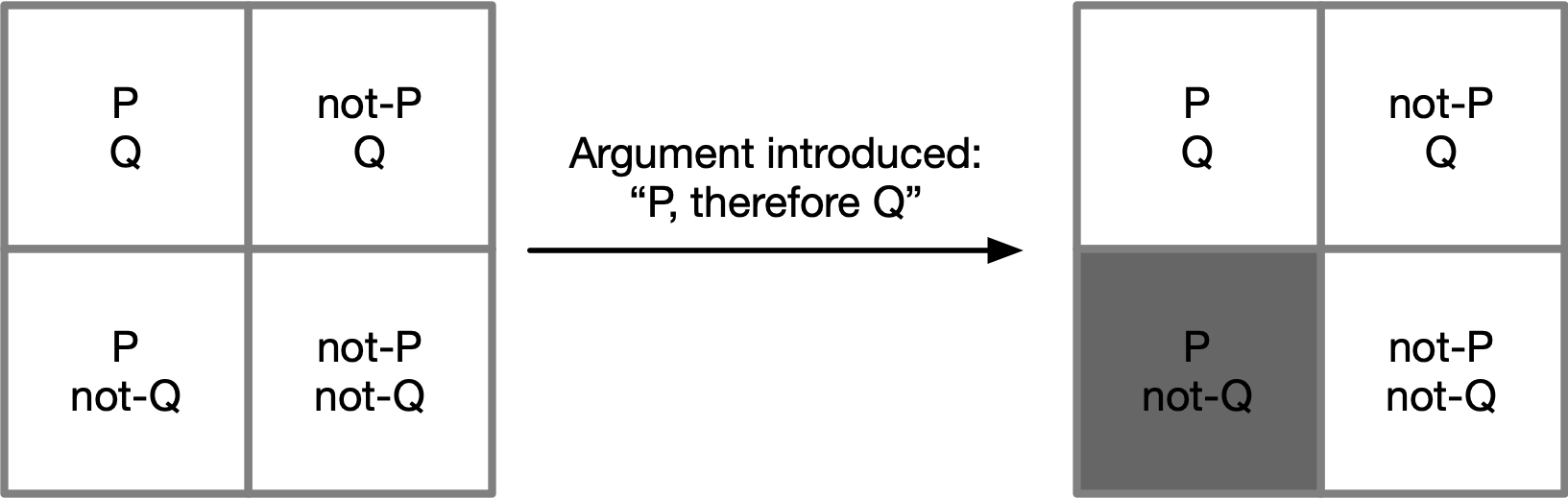}
\caption{An argument makes positions untenable in the following way:
consider two propositions, $p$ and $q$. There are four logically
possible combinations of truth values for these propositions. The
argument ``$p$, therefore $q$'' does not tell us anything about
what is the case if $p$ is false, so it leaves the not-$p$ positions
untouched. What the argument does say is that if $p$ is true, then
$q$ is true also. So it renders untenable all the positions at which
$p$ is true and $q$ is false.}
\end{figure}

\newpage

\hypertarget{description-of-strategy-space}{%
\section{Description of strategy
space}\label{description-of-strategy-space}}

\small

\begin{longtable}[]{@{}llll@{}}
\caption{Enumeration of strategies studied
\label{tab_strategies}}\tabularnewline
\toprule
Monadic/dyadic & Group & Name & Abbreviation \\
\midrule
\endfirsthead
\toprule
Monadic/dyadic & Group & Name & Abbreviation \\
\midrule
\endhead
Monadic & Bold & Self-accept, Self-reject & S+S-- \\
& & Other-accept, Other-reject & O+O-- \\
& Conservative & Self-reject, Self-Reject & S--S-- \\
& & Self-reject, Self-accept & O--O+ \\
& & Other-reject, Other-reject & O--O-- \\
& & Other-reject, Other-accept & O--O+ \\
& & Self-accept, Self-accept & S+S+ \\
& & Other-accept, Other-accept & O+O+ \\
Dyadic & Compromising & Self-reject, Other-reject & S--O-- \\
& & Self-accept, Other-reject & S+O-- \\
& & Self-accept, Other-accept & S+O+ \\
& & Other-reject, Self-reject & O--S-- \\
& & Other-accept, Self-reject & O+S-- \\
& & Other-accept, Self-accept & O+S+ \\
& Refusenik & Other-reject, Self-accept & O--S+ \\
& & Self-reject, Other-accept & S--O+ \\
& & Exit & Exit \\
\bottomrule
\end{longtable}

\normalsize

\hypertarget{comparison-with-other-strategy-spaces-studied}{%
\subsection{Comparison with other strategy spaces
studied}\label{comparison-with-other-strategy-spaces-studied}}

In (1), four strategies are studied: Convert, Fortify, Attack, and
Undercut. These strategies, like the strategies studied here, are
defined by rules for choosing premises and conlusions of arguments.

\begin{itemize}
\tightlist
\item
  The \textbf{Convert} strategy uses premises accepted by opponent and a
  conclusion accepted by proponent, thus it closely resembles
  Other-accept, Self-accept (\textbf{O+S+}) in our model.
\item
  The \textbf{Fortify} strategy uses premises accepted by proponent and
  a conclusion accepted by the proponent, thus it closely resembles
  Self-accept, Self-accept (\textbf{S+S+}) in our model.
\end{itemize}

The Attack and Undercut strategies are special cases of the Fortify and
Convert strategies, respectively, in that each uses only conclusions
that satisfy a further restriction that the opponent \emph{rejects} the
conclusion. So the number of valid arguments that can be generated by
these strategies is a proper subset of the arguments that can be
generated by Convert and Fortify. These strategies have no direct
analogue in our framework.

In Betz's model, debaters resort to random argumentation strategies when
they can no longer produce an argument using their prescribed strategy.
This is an important difference from our framework since we strictly
limit debaters to arguments consistent with their defining rule, and
permit debates to reach stalemate thereafter.

Two other differences to note. In Betz's original simulations:

\begin{enumerate}
\def\labelenumi{\arabic{enumi}.}
\item
  Turns strictly alternate between debaters. In the present work, we
  randomize who may generate an argument with each turn. This enables us
  to keep the size of the state space smaller by a factor of 2, because
  no record needs to be kept of whose turn it is.
\item
  All debates are terminated once the space of logical positions is
  reduced beyond a fixed threshold. In contrast, we allow a continuously
  variable parameter, $\delta$, to affect the expected length of the
  debate.
\end{enumerate}

\newpage

\hypertarget{detailed-methods}{%
\section{Detailed methods}\label{detailed-methods}}

In the following, we give a formal account of the model. We describe the
debaters' possible positions, their possible arguments, and how we
formalize the state of the debate at each point in time. Thereafter, we
specify among which strategies the debaters can choose to formulate
arguments. If the state space is sufficiently small, we show that the
expected dynamics of a debate can be computed exactly. To this end, we
represent the debate as a Markov chain.

\hypertarget{a-description-of-the-debating-process}{%
\subsection{A description of the debating
process}\label{a-description-of-the-debating-process}}

\hypertarget{positions}{%
\subsubsection{Positions}\label{positions}}

We consider a slightly more general model than in the main text. We
consider two debaters who hold opinions on $n$ different
\emph{propositions} $P_1$, $P_2$, \ldots, $P_n$. Each of these
propositions is either true or false. We describe a possible opinion on
all these propositions by a vector
$\mathbf{p}\!=\!(p_1,\ldots,p_n)\!\in\! \{0,1\}^n$. An entry
$p_i\!=\!1$ indicates that the respective debater considers
proposition $P_i$ to be true. Conversely, $p_i\!=\!0$ means the
debater considers $P_i$ to be false (or equivalently, the debater
considers the negation $\neg P_i$ to be true). It is sometimes useful
to have an explicit enumeration of possible views. To this end, we think
of each vector $\mathbf{p}$ as an $n$-digit binary number. That is,
we use the representations $\mathbf{p^0} \!=\! (0,0,\ldots,0,0)$,
$\mathbf{p^1}\!=\!(0,0,\ldots,0,1)$,
$\mathbf{p^2} \!=\! (0,0,\ldots,1,0)$, and so on. We refer to each
tuple $\mathbf{p^k}$ as a \emph{position}, and to the set of all such
positions as $\mathcal{P}$. The set $\mathcal{P}$ contains $2^n$
elements. Without loss of generality, we assume that the last element
(1,\ldots,1) represents the factual truth (unbeknownst to the debaters).

\hypertarget{arguments}{%
\subsubsection{Arguments}\label{arguments}}

We model a debate as a dynamical process during which the debaters put
forward arguments. An argument consists of a premise and a conclusion.
Each premise is a combination of propositions ($P_1,\ldots,P_n$) and
negations of propositions ($\neg P_1,\ldots,\neg P_n$). These are
statements that are all taken for granted for the sake of the argument.
The conclusion is a single proposition (or negation of a proposition)
that is argued to be true if the premise holds. For example, one
possible argument is \begin{equation}
\text{$A_1$: If $P_1$ is true and $P_2$ is false, then $P_3$ is false.}
\end{equation} To write this argument in a compact manner, we use the
notation \begin{equation}
P_1, \neg P_2 \vdash \neg P_3.
\end{equation} We assume arguments satisfy basic syntactic rules. First,
if some proposition $P_i$ is in the argument (either as premise or
conclusion), $\neg P_i$ cannot be in the argument, and vice versa.
This rule prevents self-contradicting arguments, such as
$P_1,P_2 \vdash \neg P_1$. Similarly, we rule out arguments in which
the same proposition $P_i$ appears twice (for example, once as a
premise, and once as a conclusion). Finally, we do not allow debaters to
make an argument that would rule out the truth. That is, we do not
permit arguments for which the set of positions that satisfy all the
premises and the negation of the conclusion is a superset of the true
position. Given that herein, we defined the truth to be the position
$(1,\ldots,1)$, this condition rules out arguments like
$P_1,P_2 \vdash \neg P_3$. We refer to arguments that satisfy the
above basic rules as \emph{valid}. In the following we assume for
simplicity that premises consist of two propositions; but the basic
framework could easily be generalized to allow for more complex
arguments.

For counting the number of possible arguments, we note that two
arguments can be syntactically different, while having the same effect
on the dialectical space of positions. For example, for $n\!=\!3$
propositions, all of the following arguments have the same effect of
ruling out position (1,1,0), \begin{equation}
\setlength{\arraycolsep}{1mm}
\begin{array}{rclcrcl}
P_1, P_2 &\vdash &P_3,
& \qquad \qquad
&P_2, P_1 &\vdash &P_3,\\
P_1, \neg P_3 &\vdash &\neg P_2, 
&
&\neg P_3, P_1 &\vdash &\neg P_2,\\
P_2, \neg P_3 &\vdash &\neg P_1, 
&
&\neg P_3, P_2 &\vdash &\neg P_1.
\end{array}
\end{equation} In the following, we identify such arguments with each
other. Two arguments $A_1$ and $A_2$ are considered equivalent if
they eliminate the same set of positions. That is, $A_1$ and $A_2$
belong to the same equivalence class if, for every position $p \in P$,
the position is ruled out by $A_1$ if and only if it is ruled out by
$A_2$. We denote this by $[A_1] \!=\! [A_2]$. Here, $[A]$
represents the equivalence class containing all arguments that have the
same eliminative effect on the debate. For $n$ propositions and all
arguments having two premises, there are $8n(n\!-\!1)(n\!-\!2)/6$ such
equivalence classes (the factor $8\!=\!2^3$ captures that for each of
the three propositions, we also need to consider their negations; the
factor 6 captures the number of elements in each class).

\hypertarget{tenable-positions}{%
\subsubsection{Tenable positions}\label{tenable-positions}}

Over the course of a debate, arguments may rule out certain positions.
We say a position is \emph{tenable} if it has not been ruled out yet by
any of the previously given arguments. Accordingly, let $T(t)$
describe the set of all tenable positions after the first $t$
arguments have been made. It follows that the set of possible tenable
positions is non-increasing, $T(t\!+\!1) \!\subseteq\! T(t)$ for all
$t\!\ge\!0$. Initially, we set $T(0)\!=\!\mathcal{P}$, the set of
all possible positions. As an example, consider a debate with three
propositions, and suppose a debater uses the argument $A_1$ defined
above, $P_1, \neg P_2 \vdash \neg P_3$. For that debate, we obtain
\begin{equation}
\begin{array}{l}
T(0) = \big\{(0,0,0), (0,0,1), (0,1,0),  (0,1,1), (1,0,0), (1,0,1), (1,1,0), (1,1,1)\big\},\\[0.1cm]
T(1) = \big\{(0,0,0), (0,0,1), (0,1,0), (0,1,1), (1,0,0), (1,1,0), (1,1,1)\big\}.
\end{array}
\end{equation} That is, argument $A_1$ eliminates the position
$(1,0,1)$ from the set of tenable positions. In principle, any subset
of $\mathcal{P}$ that contains the truth $(1,\ldots,1)$ is a
possible set of tenable positions that could arise during a debate. We
refer to the set of all such subsets as $\mathcal{T}$. In total,
$\mathcal{T}$ has $2^{2^{n}-1}$ elements (it corresponds to the
power set of all $2^n\!-\!1$ positions in
$\mathcal{P}\!\setminus\!\big\{(1,\ldots,1)\big\}$, with the truth
$(1,\ldots,1)$ then being added to each element). For example, if
there are only two propositions, the respective set has eight elements,
\begin{equation} \setlength{\arraycolsep}{0.05cm}
\begin{array}{rcl}
\mathcal{T} &= &\Big\{\,\big\{(1,1)\big\}, \big\{(0,0),(1,1)\big\}, \big\{(0,1),(1,1)\big\}, \big\{(1,0),(1,1)\big\}, \big\{(0,0), (0,1), (1,1)\big\},\ldots\\[0.1cm]
&&   \big\{(0,0), (1,0), (1,1)\big\}, \big\{(0,1), (1,0), (1,1)\big\}, \big\{(0,0), (0,1), (1,0), (1,1)\big\}\Big\}
\end{array}
\end{equation} As illustrated in the main text, the set of possible
positions can be represented by a hypercube. There, a corner of the
hypercube is colored black if the respective position is tenable, and it
is colored as grey otherwise. The corner $(1,\ldots,1)$ is always
colored black.

\hypertarget{a-model-of-the-debating-process}{%
\subsubsection{A model of the debating
process}\label{a-model-of-the-debating-process}}

The state of a debate after the exchange of $t$ arguments can be
defined as a triplet
$\omega(t) \!=\! \big(T(t),\, \mathbf{p'}(t),\, \mathbf{p''}(t)\big)$;
we sometimes drop the time-dependence and write
$\omega \!=\! \big(T,\, \mathbf{p'},\, \mathbf{p''}\big)$. The first
entry $T\!\in\!\mathcal{T}$ describes the set of positions that are
still tenable. The entries $\mathbf{p'}, \mathbf{p''}$ describe the
two debaters' current positions. In particular, both positions need to
be tenable, $\mathbf{p'},\mathbf{p''} \!\in\! T$. We refer to the set
of all such states $\omega$ as $\Omega$.

At the beginning, we assume the players' initial positions
$\mathbf{p'}(0)$ and $\mathbf{p''}(0)$ are independently chosen
uniformly at random from all positions in $T(0)\!=\!\mathcal{P}$. Over
the course of a debate, the two debaters exchange arguments. To this
end, in each round, one of the two debaters is chosen at random. This
debater is then permitted to put forward an argument (which might depend
on the debater's strategy, as defined further below). If there is no
valid argument consistent with the debater's strategy, the debater
passes and the state $\omega$ does not change. If the debater produces
a valid argument, the set of tenable positions $T$ may decrease. As a
result, a debater's current position may become untenable. In that case,
we assume the debater moves to the closest position that is still
tenable. If there are multiple tenable positions that are equally close,
one of them is chosen randomly. To quantify the closeness of two
positions $\mathbf{p}$ and $\mathbf{\tilde p}$, we use the Hamming
distance -- which is equivalent to counting the number of propositions
on which the two positions disagree: \begin{equation}
\lVert \mathbf{p} - \mathbf{\tilde p} \rVert = \sum_{i=1}^n \lvert p_i \!-\! \tilde p_i \rvert.
\end{equation}

\noindent After each argument, the debate ends with probability
$\delta$. In particular, the debate's final round $\tau$ is a random
variable with a geometric distribution. A realization of a debate is a
sequence of states
$\vect{\sigma} \!=\! \big(\omega(0), \ldots, \omega(\tau)\big)$, such
that each entry is a possible successor of the previous entry given the
rules of the game outlined above.

\hypertarget{the-debaters-payoffs}{%
\subsubsection{The debaters' payoffs}\label{the-debaters-payoffs}}

We assume debaters have two objectives, accuracy and obstinacy. Accuracy
means debaters prefer to be close to the truth eventually. Obstinacy
means they prefer to change their position as rarely as possible.

To quantify these objectives, consider a particular realization of a
debate,
$\vect{\sigma} \!=\! \big(\omega(0), \ldots, \omega(\tau)\big)$. If a
debater's last position is $\mathbf{p}(\tau)$, we define the debater's
accuracy as \begin{equation} \label{Eq:accuracy}
v_\vect{\sigma} = n - \big\lVert\, \mathbf{p}(\tau) \!-\! (1,\ldots,1) \,\big\rVert = \big\lVert\, \mathbf{p}(\tau) \,\big\rVert  .
\end{equation} This accuracy is a number between 0 and $n$. It
reflects how many of the debater's views are correct by the time the
debate ends. A debater's obstinacy penalty $d_\vect{\sigma}$ is the
aggregated number of changes in the debater's position (for each
proposition) over the course of the debate. Formally, we define it as
the sum of the hamming distances, \begin{equation}
d_\vect{\sigma} = \sum_{t=1}^\tau \big\lVert \mathbf{p}(t)\!-\!\mathbf{p}(t\!-\!1) \big\rVert.
\end{equation} Taken together, we define the debater's payoff as
\begin{equation} \label{eq:PayoffRealization}
u_\vect{\sigma} =w\!\cdot\!v_\vect{\sigma} - d_\vect{\sigma}.
\end{equation} Here, $w$ reflects the relative weight the debater
places on being accurate.

\hypertarget{a-description-of-the-players-strategies}{%
\subsection{A description of the players'
strategies}\label{a-description-of-the-players-strategies}}

\hypertarget{strategies}{%
\subsubsection{Strategies}\label{strategies}}

When it is their turn, debaters need to decide which arguments to make.
To do so, each debater employs a \emph{strategy}. A strategy is a rule
that tells the debater which argument to choose, at each possible
occasion. In general, this choice might depend on all previously made
arguments, and on the two debaters' positions. In the following, we
assume strategies only depend on the debaters' current positions. More
specifically, we consider strategies that can be represented by a pair
$\mathbf{q}\!=\!(q_1,q_2)$. The first component $q_1$ encodes what
kind of propositions are used as the premise of the argument. We allow
four possibilities: (\emph{i}) When $q_1\!=\!S^+$, the premise set
only contains propositions the proponent accepts. (\emph{ii}) When
$q_1\!=\!S^-$, the premise set contains at least one proposition the
proponent rejects. (\emph{iii}) When $q_1\!=\!O^+$, the premise set
only contains propositions the opponent accepts. (\emph{iv}) When
$q_1\!=\!O^-$, the premise set contains at least one proposition that
the opponent rejects. Analogously, $q_2\!\in\!\{S^+,S^-,O^+,O^-\}$
encodes what kinds of propositions are used as the conclusion of the
argument, with the added restriction that the conclusion set is always a
singleton.

We call an argument ``feasible'' for a given proponent if it is valid
and if it is consistent with the proponent's strategy. We call an
equivalence class of arguments feasible if it contains at least one
feasible argument. When during the debate, there are several feasible
classes of arguments, we assume the proponent chooses one of them
uniformly at random. When there is no feasible class of arguments, the
proponent passes. Such a case could occur, for example, if the
proponent's current position is the truth, (1,\ldots,1) and the
proponent's strategy is ($S^+,S^-$). All arguments consistent with
that strategy would refute the truth; hence, they are invalid by
definition. We give an example of a possible debating dynamics in
\textbf{Example 1}.

We note that the choice of a strategy affects how likely debaters are to
change their positions over time. As an example, consider again the
strategy ($S^+, S^-$). Debaters with that strategy aim to falsify
their own position: they take a premise they accept while trying to
argue for a conclusion they disagree with. As long as the resulting
argument is valid (e.g., it does not rule out the truth), it makes the
proponent's position untenable. Thus, such a strategy potentially
induces the debater to frequently change positions over time. At the
same time, it may be an effective means to eventually reach the true
position.

In addition to the sixteen strategies of the form
$\mathbf{q}\!=\!(q_1,q_2)$, we allow for an additional strategy called
`Exit'.\\
This strategy may be interpreted as a complete refusal to engage in the
debate. If one of the debaters employs this strategy, the debate ends
immediately; the final state of the debate is equal to the initial
state.

\begin{table}[t]
\centering
\small
\fbox{
\parbox[t]{\textwidth}{
\begin{Example} 
As an example of a possible debating dynamics, consider a debate with three propositions. 
Suppose the current state is $\omega\!=\!(T,\mathbf{p'},\mathbf{p''})$, where the set of tenable positions is
$$T\! =\!  \big\{(0,1,0), (0,1,1), (1,0,0), (1,1,0), (1,1,1)\big\},$$ 
and the players' positions are $\mathbf{p'}\!=\!(0,1,1)$ and $\mathbf{p''}\!=\!(1,1,1)$. 
Moreover, suppose the first debater is randomly chosen to make the next argument, and this debater uses strategy $\mathbf{q'}\!=\!(S^+,O^-)$.
Given the first debater's position and strategy, a possible premise may consist of the propositions $\neg P_1$ and $P_2$. 
Similarly, given the opponent's position, $\neg P_3$ is a possible conclusion. 
Since the respective argument $\neg P_1, P_2 \vdash \neg P_3$ does not reject the truth, it is feasible.  
If the first debater makes that argument, it rules out the position (0,1,1), which happens to be the debater's own position. 
This debater then randomly switches to a close-by position, such as the factually true position (1,1,1), which is a Hamming distance of 1 away from the original position. 
In that case, the next state is
\begin{equation} \label{Eq:ExampleState}
\tilde T\! =\!  \big\{(0,1,0), (1,0,0), (1,1,0), (1,1,1)\big\},     \mathbf{\tilde p'}\!=\mathbf{\tilde p''}\!=\!(1,1,1).
\end{equation}
\end{Example}
}}
 \caption{} \label{Table:Example1}
\end{table}

\hypertarget{payoffs-of-strategies}{%
\subsubsection{Payoffs of strategies}\label{payoffs-of-strategies}}

When two debaters with strategies $\mathbf{q'}$ and $\mathbf{q''}$
interact, they naturally induce a probability distribution
$\vect{x}(\mathbf{q'},\mathbf{q''})\!=\!(x_\vect{\sigma})$ over the
space of all possible sequences
$\vect{\sigma}\!=\!\big(\omega(0), \ldots, \omega(\tau)\big)$. Here,
each entry $x_\vect{\sigma}$ is the likelihood of observing the
realized debate $\vect{\sigma}$, depending on the players' debating
strategies. We define the player's overall payoffs as their expected
payoff, \begin{equation} \label{Eq:ExpectedPayoff}
u(\mathbf{q'},\mathbf{q''}) = \mathbb{E}_\vect{\sigma} [ u_\vect{\sigma}] = \sum_{\vect{\sigma}} x_\vect{\sigma} \cdot u_\vect{\sigma}.
\end{equation} That is, the payoff is the weighted average over all
possible realized payoffs $u_\vect{\sigma}$ as defined by
\eqref{eq:PayoffRealization}, weighted by how likely that realization is
observed given the debaters' strategies. Given the above rules and the
debaters' strategies, it is straightforward to approximate the resulting
payoffs with simulations. To this end, it suffices to generate
sufficiently many realizations
$\vect{\sigma}\!=\!\big(\omega(0), \ldots, \omega(\tau)\big)$, and to
compute the average of the realized payoffs. We provide an
implementation of such a simulation in our code repository.

\hypertarget{a-markov-chain-representation-of-the-debate}{%
\subsection{A Markov chain representation of the
debate}\label{a-markov-chain-representation-of-the-debate}}

In the following, we describe a method that allows an explicit
computation of payoffs when the number of positions is sufficiently
small (for the results in the main text, $n\!=\!3$). We note that if
one of the two debaters adopts the Exit strategy, the payoffs are
straightforward to compute. In that case, both of the debaters obtain a
payoff of $u\!=\! w \times \nicefrac{n}{2}$ (a priori, both of the
debaters are correct on any single proposition with probability
$\nicefrac{1}{2}$). In the following, we thus assume that neither
player adopts the Exit strategy.

\hypertarget{computing-the-debaters-expected-accuracy}{%
\subsubsection{Computing the debaters' expected
accuracy}\label{computing-the-debaters-expected-accuracy}}

To compute payoffs, we represent the dynamics of the debate as a Markov
chain. The possible states of the Markov chain are given by the possible
states of the debate, $\Omega\!=\!\{\omega^1,\ldots,\omega^m\}$. Given
the current state $\omega^i$ and the debaters' strategies
$\mathbf{q'}$ and $\mathbf{q''}$, we can compute the transition
probability $M_{ij}$ that after the next argument, the state is
$\omega^j$.

To this end, let $\mathcal{A}\!=\!\mathcal{A}(\omega^i,\mathbf{q})$
denote the set of feasible classes of arguments that a debater with
strategy $\mathbf{q}$ could produce in state $\omega^i$. To refer to
these sets for the two respective players, we write
$\mathcal{A'}\!=\!\mathcal{A}(\omega^i,\mathbf{q'})$ and
$\mathcal{A''}\!=\!\mathcal{A''}(\omega^i,\mathbf{q''})$. We use
$|\mathcal{A}|$ to refer to the number of elements in $\mathcal{A}$
(i.e., the number of distinct arguments a debater could produce). For a
given class $[A]$ of arguments, let
$\mathbb{P}'\big([A]\,\big|\,\omega^i,\mathbf{q'}\big)$ denote the
probability that an argument in that class would be produced by the
first debater with strategy $\mathbf{q'}$ in state $\omega^i$.
Because arguments are chosen uniformly at random across all feasible
equivalence classes, \begin{equation}
\mathbb{P'}\big([A]\,\big|\,\omega^i,\mathbf{q'}\big)\!=\!\left\{
\begin{array}{ll}
1/|\mathcal{A'}(\omega^i,\mathbf{q'})|  &\text{if } [A]\!\in\!\mathcal{A'}(\omega^i,\mathbf{q'})\\
0   &\text{otherwise}.
\end{array}
\right.
\end{equation} An analogous definition applies to the respective
quantity $\mathbb{P}''\big([A]\,\big|\,\omega^i,\mathbf{q''}\big)$ for
the second debater. Given an argument in class $[A]$ has been
produced, we define an indicator function with
$e\big(\omega^j \,\big|\, \omega^i, [A]\big)$ that indicates whether
or not the argument can induce the state to change from $\omega^i$ to
$\omega^j$, \begin{equation} \label{Eq:indicator}
e\big(\omega^j \,\big|\, \omega^i, [A]\big)\!=\!\left\{
\begin{array}{ll}
1   &\text{if the state } \omega^j \text{ can be reached from } \omega^i \text{ after an argument in } [A]\\
0   &\text{otherwise.}
\end{array}
\right.
\end{equation} Based on this notation, we define the combined
probability that when the first debater with strategy $\mathbf{q'}$ is
to make an argument in state $\omega^i$, the argument will be in class
$[A]$ leading to the new state $\omega^j$, \begin{equation}
\mathbb{P'}\big(\omega^j, [A]\,\big|\,\omega^i,\mathbf{q'}\big)\!=\!\left\{
\begin{array}{cl}
\displaystyle \frac{e\big(\omega^j \big| \omega^i,[A]\big)\cdot \mathbb{P}'\big([A]\big|\omega^i,\mathbf{q'}\big)}{\sum_\omega e\big(\omega\,\big|\, \omega^i,[A]\big)\cdot \mathbb{P'}\big([A]\big|\omega^i,\mathbf{q'}\big)}  &\text{if } [A]\!\in\!\mathcal{A'}(\omega^i,\mathbf{q'})\\[0.5cm]
0   &\text{otherwise}
\end{array}
\right.
\end{equation} Again, an analogous definition applies to the second
debater's probability
$\mathbb{P''}\big(\omega^j, [A]\,\big|\,\omega^i,\mathbf{q''}\big)$.
After these preparations, we can formally define the Markov chain's
transition probability from state $\omega^i$ to
$\omega^j\! \neq \!\omega^i$ as \begin{equation} \label{Eq:Mij}
M_{ij} =\frac{1}{2} \sum_{[A]} \mathbb{P'}\big(\omega^j, [A]\,\big|\,\omega^i,\mathbf{q'}\big)
+
\frac{1}{2} \sum_{[A]} \mathbb{P''}\big(\omega^j, [A]\,\big|\,\omega^i,\mathbf{q''}\big)
\end{equation} Here, the factor $1/2$ represents that each of the two
debaters has the same chance to make the next argument. In
\textbf{Example 2}, we illustrate the computations
necessary to derive such a transition probability $M_{ij}$ for one
particular case. For the remaining transition probability $M_{ii}$, we
obtain \begin{equation}
\textstyle M_{ii} = 1- \sum_{j\neq i} M_{ij}.
\end{equation} This probability covers the cases that either a debater
was not able to produce a feasible argument, or that a debater made an
argument that led to no change (possibly because it was already made
before).

We can collect these probabilities to move from state $\omega^i$ to
$\omega^j$ in an $m \times m$ transition matrix $M\!=\!(M_{ij})$.
Let $\mathbf{y}(t) \!=\! \big(y_i(t)\big)$ be the probability
distribution that reflects how likely players are in state $\omega^i$
at time $t$. For the initial probability distribution
$\mathbf{y}(0)$, we consider all states for which the set of tenable
positions coincides with the set of all positions. There are $2^{2n}$
such states (they only differ in the initial positions assigned to the
two debaters). Hence, it follows that \begin{equation}
y_i(0) = \left\{
\begin{array}{ll}
\frac{1}{2^{2n}}    &\text{if } \omega^i=(T,\mathbf{p'},\mathbf{p''}),   \text{with } T\!=\!\mathcal{P} \text{ and }  \mathbf{p'},\mathbf{p''}  \text{ arbitrary}\\
0   &\text{otherwise}.
\end{array}
\right.
\end{equation} For all subsequent probability distributions, it follows
from the theory of Markov chains that
$\mathbf{y}(t)=\mathbf{y}(0)M^t$. In particular, because the length of
the debate is geometrically distributed, we can compute the expected
final state of the debate as \begin{equation}
\mathbf{y}:= \sum_{t=0}^\infty \delta^t (1\!-\!\delta) \mathbf{y}(t) = (1\!-\!\delta)\mathbf{y}(0) \sum_{t=0}^\infty \delta^t M^t = (1\!-\!\delta)\mathbf{y}(0)(I_m-\delta M)^{-1}. 
\end{equation} Here, $I_m$ is the $m\!\times\!m$ identity matrix,
and $(I_m-\delta\!\cdot \! M)^{-1}$ refers to the respective inverse
matrix (which is guaranteed to exist for all $\delta\!<\!1$). Based on
this final state, we can compute the debater's expected accuracy $v$.
To this end, we take the probability $y_\omega$ that a given state
$\omega\!=\!(T,\mathbf{p'},\mathbf{p''})$ is the final state of the
debate, times the accuracy of the debater's position, according to Eq.
\eqref{Eq:accuracy}, $\lVert \mathbf{p'} \rVert$. By summing up, we
obtain the following formulas for the expected accuracies of the two
players, \begin{equation}
v' = \sum_{\omega \in \Omega} y_\omega \cdot \lVert \mathbf{p'} \rVert,
\quad \qquad
v'' = \sum_{\omega \in \Omega} y_\omega \cdot \lVert \mathbf{p''} \rVert. 
\end{equation}

\hypertarget{computing-the-debaters-expected-obstinacy-penalty}{%
\subsubsection{Computing the debaters' expected obstinacy
penalty}\label{computing-the-debaters-expected-obstinacy-penalty}}

To compute the expected obstinacy penalty, we first compute a vector
$\mathbf{z}\!=\!(z_\omega)$ that indicates how often we visit state
$\omega$ on average during the course of a debate. By summing over all
possible time steps in which we might visit the respective state, we
obtain \begin{equation}
\mathbf{z}:= \sum_{t=0}^\infty \delta^t \mathbf{y}(t) = \mathbf{y}(0) \sum_{t=0}^\infty \delta^t M^t = \mathbf{y}(0)(I_m-\delta M)^{-1}. 
\end{equation} In addition, we define matrices $Q'\!=\!(Q'_{ij})$ and
$Q''\!=\!(Q''_{ij})$ that record the hamming distance for the two
debaters, respectively, between state
$\omega^i\!=\!(T,\mathbf{p'},\mathbf{p''})$ and
$\omega^j\!=\!(\tilde T, \mathbf{\tilde p'},\mathbf{\tilde p''})$. We
obtain \begin{equation}
Q'_{ij} = \lVert \mathbf{\tilde p'} \!-\! \mathbf{p'} \rVert \qquad \text{and} \qquad  Q''_{ij} = \lVert \mathbf{\tilde p''} \!-\! \mathbf{p''} \rVert. 
\end{equation} To compute the debater's expected obstinacy penalty
$d$, we multiply the expected number $z_{\omega^i}$ of visits to
each state $\omega^i$ by the cost $Q_{ij}$ of the subsequent
transition to state $\omega^j$. This cost needs to be weighted by the
probability $\delta M_{ij}$ of the respective transition happening
($\delta$ is the probability that there is another transition at all).
By summing up over all possible initial states $\omega^i$ and
subsequent states $\omega^j$, we obtain \begin{equation}
d' =  \delta \sum_{i,j} z_{\omega^i} \cdot M_{ij} \cdot Q'_{ij}
\qquad \text{and} \qquad
d'' =  \delta \sum_{i,j} z_{\omega^i} \cdot M_{ij} \cdot Q''_{ij}.
\end{equation} Based on the above formulas for the players' accuracy and
their obstinacy penalty, we can explicitly compute the players' payoffs
according to Eq. \eqref{eq:PayoffRealization}. Again, we provide an
implementation of this payoff computation in our code repository. We
have checked the implementation's accuracy by comparing its results to
simulations.

\newpage 

\begin{table}[H]
\centering
\footnotesize \setstretch{1.0}
\fbox{
\parbox[t]{\textwidth}{
\begin{Example} We illustrate how to compute transition probabilities based on Eq. \eqref{Eq:Mij} by discussing a special case. Following up on \textbf{Example \ref{Table:Example1}}, the current state $\omega^i\!=\!(T,\mathbf{p'},\mathbf{p''})$ is again
\begin{equation}
T\! =\!  \big\{(0,1,0), (0,1,1), (1,0,0), (1,1,0), (1,1,1)\big\}, \quad \mathbf{p'}\!=\!(0,1,1), \quad \mathbf{p''}\!=\!(1,1,1).
\end{equation}
Moreover, the debaters' strategies are $\mathbf{q'}\!=\!(S^+,O^-)$ and $\mathbf{q''}\!=\!(S^+,S^-)$.
We ask how likely it is to make the transition towards state $\omega^j\!=\!(\tilde T, \mathbf{\tilde p'}, \mathbf{\tilde p''})$ given by
\begin{equation}
\tilde T\! =\!  \big\{(0,1,0), (1,0,0), (1,1,0), (1,1,1)\big\},\quad \mathbf{\tilde p'}\!=\mathbf{\tilde p''}\!=\!(1,1,1).
\end{equation}
That is, we ask how likely the position $(0,0,1)$ becomes untenable, and the first debater moves to $(1,1,1)$ as a result. 
To this end, we consider two cases, depending on which debater is chosen to make the next argument. 

\begin{description}
\item[\underline{Case 1: First debater makes next argument.}] Given the current positions $\mathbf{p'}$ and $\mathbf{p''}$ of the two debaters, the following arguments are consistent with the first debater's strategy,
$$
\setlength{\arraycolsep}{1mm}
\begin{array}{rclcrcl}
\neg P_1, P_2 &\vdash &\neg P_3,
& \qquad \qquad
&P_2, \neg P_1 &\vdash &\neg P_3,\\
\neg P_1, P_3 &\vdash &\neg P_2, 
&
&P_3, \neg P_1 &\vdash &\neg P_2,\\
P_2, P_3 &\vdash &\neg P_1, 
&
&P_3, P_2 &\vdash &\neg P_1.
\end{array}
$$
We note that the two arguments in the last row rule out the truth (1,1,1). 
Hence they are infeasible. 
The other four arguments all rule out the same position (0,1,1), and hence they are in the same equivalence class $[A^*]$, with $A^*:= (\neg P_1,P_2\vdash \neg P_3)$. That is, we obtain 
\begin{equation} \label{Eq:Example_2a}
\mathbb{P}'\big([A^*] \,\big|\, \omega^i,\mathbf{q'}\big) = 1.
\end{equation}
As the next step, we need to identify the possible next states, once the argument $A^*$ has been made.
Given $A^*$, it follows that $\tilde T$ is the next set of tenable states. However, given that the first debater's position has been removed, there are two possibilities for what the debater's next position could be, (0,1,0) and (1,1,1) [both have Hamming distance one to the debater's original position; the remaining position with Hamming distance one, (0,0,1), is not tenable]. This means there are two possible states that might succeed $\omega^i$ after argument $A^*$: the intended target state $\omega^j$ as defined above, and an alternative state $\omega^k$ that is identical to $\omega^j$ except that $\mathbf{p'}\!=\!(0,1,0)$.
Using our notation in Eq. \eqref{Eq:indicator}, we get 
\begin{equation} \label{Eq:Example_2b}
e\big(\omega^j\,\big|\, \omega^i,[A^*]\big) \!=\! e\big(\omega^k\,\big|\, \omega^i,[A^*]\big) \!=\! 1 \quad \text{and} \qquad e\big(\omega\,\big|\, \omega^i,[A^*]\big)\! =\!0   \text{for all} \omega \notin \{\omega^i, \omega^j, \omega^k\}
\end{equation}
By combining Eqs. \eqref{Eq:Example_2a} and \eqref{Eq:Example_2b}, we obtain the overall probability that debater would choose an argument in class $[A^*]$ and end up in the desired state $\omega^j$ as
\begin{equation}
\mathbb{P'}\big(\omega^j, [A^*]\,\big|\,\omega^i,\mathbf{q'}\big)\!=\!\nicefrac{1}{2}. 
\end{equation}

\item[\underline{Case 2: Second debater makes next argument.}] Given the second debater's strategy and current position, this debater is unable to produce a feasible argument. Any argument would refute the debater's current position, which happens to be the true state. Any such argument is invalid by definition. In particular, this debater cannot produce an argument that would reduce the set of tenable positions. Hence,
\begin{equation}
\mathbb{P''}\big(\omega^j, [A^*]\,\big|\,\omega^i,\mathbf{q''}\big)\!=\!0. 
\end{equation} 
\end{description}
Combining these two cases, we use Eq. \eqref{Eq:Mij} to compute the overall transition probability from state $\omega^i$ to $\omega^j$,
\begin{equation}
M_{ij} = \frac{1}{2}\cdot \frac{1}{2} + \frac{1}{2}\cdot 0 = \frac{1}{4}. 
\end{equation}
\end{Example}
}}
 \caption{} \label{Table:Example2}
\end{table}
\newpage

\hypertarget{an-extension-to-the-debate-model-debater-error}{%
\subsection{An extension to the debate model: debater
error}\label{an-extension-to-the-debate-model-debater-error}}

In the model described above, debaters always choose arguments that are
valid, consistent with their strategy, and feasible given the current
state of the debate. This represents ideal debaters who never make
mistakes in their argumentation. However, real-world debaters may
occasionally employ invalid arguments, either due to logical errors,
misunderstandings, or strategic attempts to confuse the debate.

To capture this aspect of real-world debates, we extend our model to
incorporate the possibility of debaters using invalid arguments. We
introduce a parameter $\alpha \in [0,1]$ that represents the
probability that a debater employs an invalid argument when no valid
arguments are available, or the weight given to invalid arguments when
both valid and invalid arguments exist. As we model it, when a debater
makes an invalid argument, the truth is eliminated from the set of
tenable positions. We can still measure epistemic performance in the
same way as previously: the number of true beliefs a debater has. But in
such a world, the debaters will be doomed never to achieve the complete
truth, because they have rejected the corresponding position.

\hypertarget{noisy-transitions}{%
\subsubsection{Noisy transitions}\label{noisy-transitions}}

We modify the transition probability calculations to account for both
valid and invalid arguments. For a debater with strategy $\mathbf{q}$,
we define $\mathcal{A}^{\text{valid}}(\omega^i,\mathbf{q})$ as the set
of feasible and valid argument classes, and
$\mathcal{A}^{\text{invalid}}(\omega^i,\mathbf{q})$ as the set of
arguments that are consistent with the debater's strategy but invalid
(for instance, because they would rule out the truth).

Given the current state $\omega^i$, we compute the transition
probabilities as follows:

\begin{description}
\item[\underline{Case 1: Valid arguments exist, invalid arguments do not exist.}] The debater selects a valid argument uniformly at random from $\mathcal{A}^{\text{valid}}(\omega^i,\mathbf{q})$. The transition probabilities are the same as in the original model:
\begin{equation}
\mathbb{P}_{\text{noisy}}(\omega^j|\omega^i,\mathbf{q}) = \sum_{[A] \in \mathcal{A}^{\text{valid}}} \mathbb{P}(\omega^j,[A]|\omega^i,\mathbf{q})
\end{equation}

\item[\underline{Case 2: Both valid and invalid arguments exist.}] The debater selects from valid arguments with full weight, and from invalid arguments with weight $\alpha$. These contributions are then normalized:
\begin{equation}
\mathbb{P}_{\text{noisy}}(\omega^j|\omega^i,\mathbf{q}) = \frac{1}{Z} \left[ \sum_{[A] \in \mathcal{A}^{\text{valid}}} \mathbb{P}(\omega^j,[A]|\omega^i,\mathbf{q}) + \alpha \cdot \sum_{[A] \in \mathcal{A}^{\text{invalid}}} \mathbb{P}(\omega^j,[A]|\omega^i,\mathbf{q}) \right]
\end{equation}
where $Z$ is a normalization constant ensuring that probabilities sum to 1.

\item[\underline{Case 3: No valid arguments, but invalid arguments exist.}] With probability $1-\alpha$, the debater passes and the state remains unchanged. With probability $\alpha$, the debater selects an invalid argument uniformly at random:
\begin{equation}
\mathbb{P}_{\text{noisy}}(\omega^j|\omega^i,\mathbf{q}) = (1-\alpha) \cdot \delta_{ij} + \alpha \cdot \sum_{[A] \in \mathcal{A}^{\text{invalid}}} \mathbb{P}(\omega^j,[A]|\omega^i,\mathbf{q})
\end{equation}
where $\delta_{ij}$ is the Kronecker delta, equal to 1 if $i=j$ and 0 otherwise.

\end{description}

These modified transition probabilities are then used to construct the
Markov transition matrix $M$ in the same manner as described
previously. The resulting Markov process captures not only the strategic
choices of debaters but also their potential errors or strategic
violations of debate norms.

\newpage

\hypertarget{supplementary-results}{%
\section{Supplementary results}\label{supplementary-results}}

\hypertarget{impact-of-different-strategy-profiles-on-consensus}{%
\subsection{Impact of different strategy profiles on
consensus}\label{impact-of-different-strategy-profiles-on-consensus}}

\begin{figure}
\centering
\includegraphics[width=0.8\textwidth,height=\textheight]{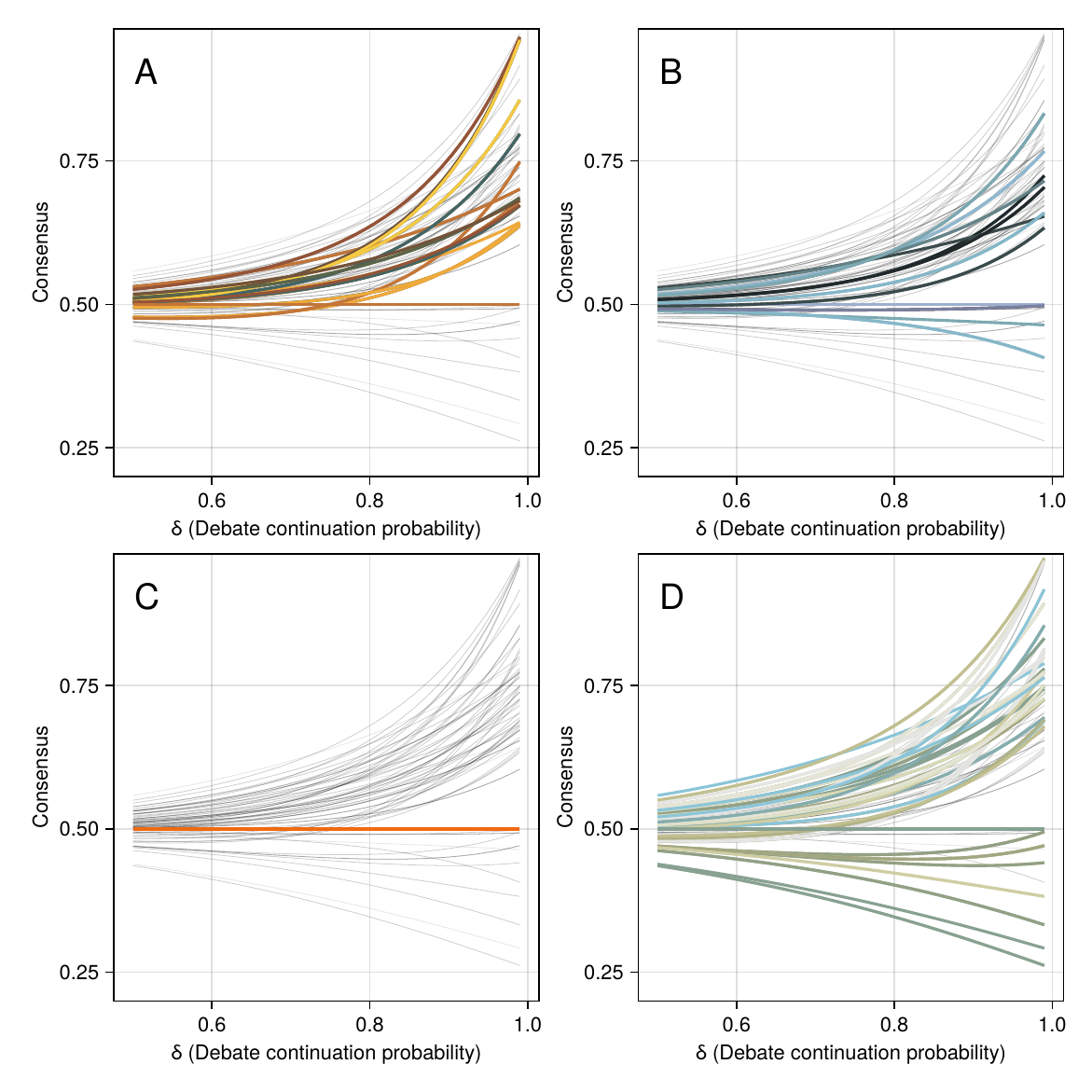}
\caption{Expected consensus over debate length, for all strategy
profiles, grouped by type. \textbf{A:} $\geq 1$ bold strategy \textbf{B:} both
conservative \textbf{C:} both refusenik \textbf{D:} $\geq 1$
consensus-building, none bold. $n$ = 3. \label{all-consensus}}
\end{figure}

\newpage

\hypertarget{impact-of-changing-size-of-the-debate-on-equilibria}{%
\subsection{Impact of changing size of the debate on
equilibria}\label{impact-of-changing-size-of-the-debate-on-equilibria}}

\begin{figure}
\centering
\includegraphics[width=1\textwidth,height=\textheight]{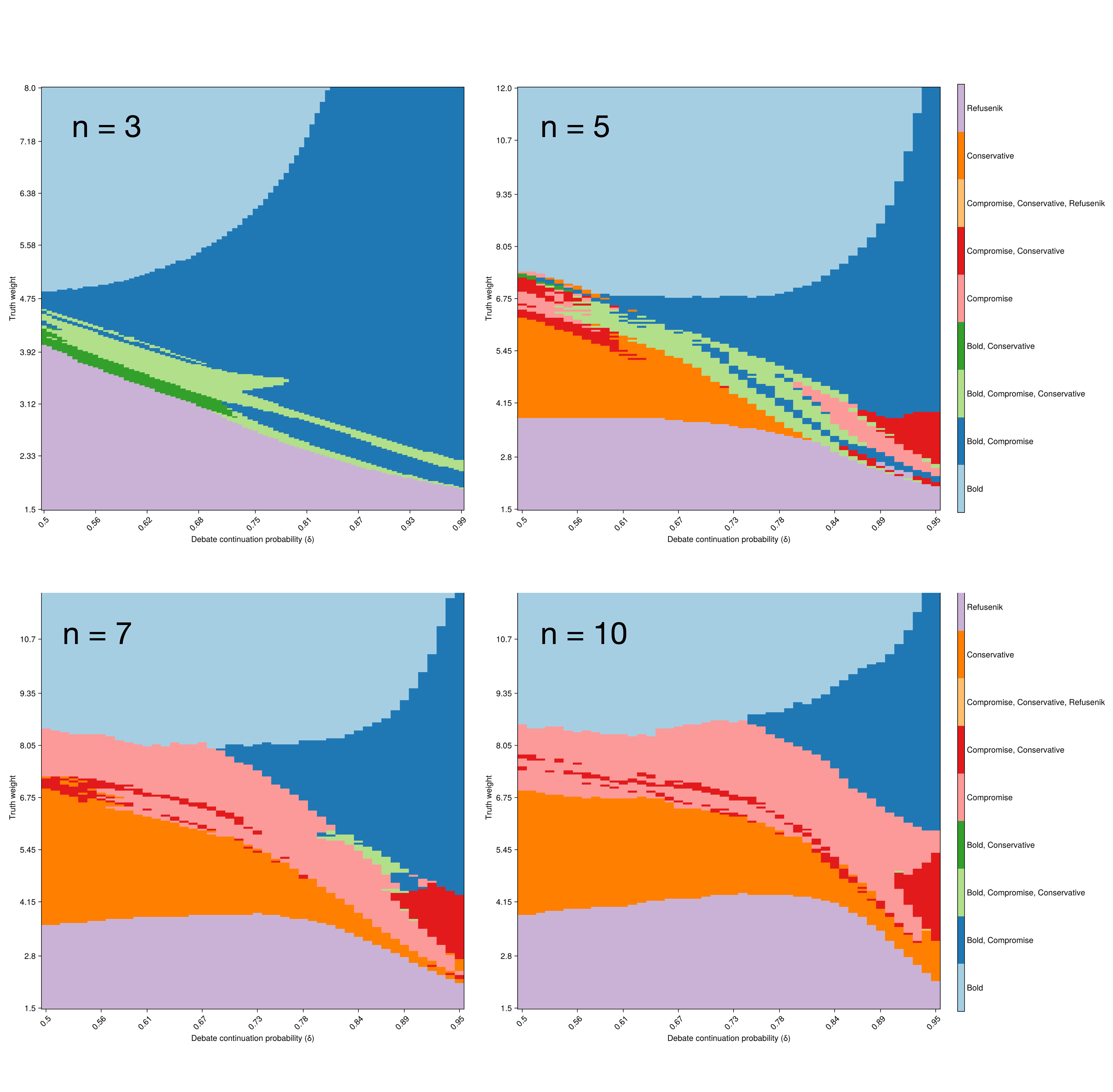}
\caption{Types of strategy in the truth-optimal equilibrium, for various $n$. For $n = 3$, results
are derived from a Markov process calculation. For larger $n$, results
are based on a simulation, with at least 200,000 trials per strategy profile
and delta value.}
\end{figure}

\newpage

\hypertarget{alternative-equilibrium-selection-criterion}{%
\subsection{Alternative equilibrium selection
criterion}\label{alternative-equilibrium-selection-criterion}}

\begin{figure}
\centering
\includegraphics{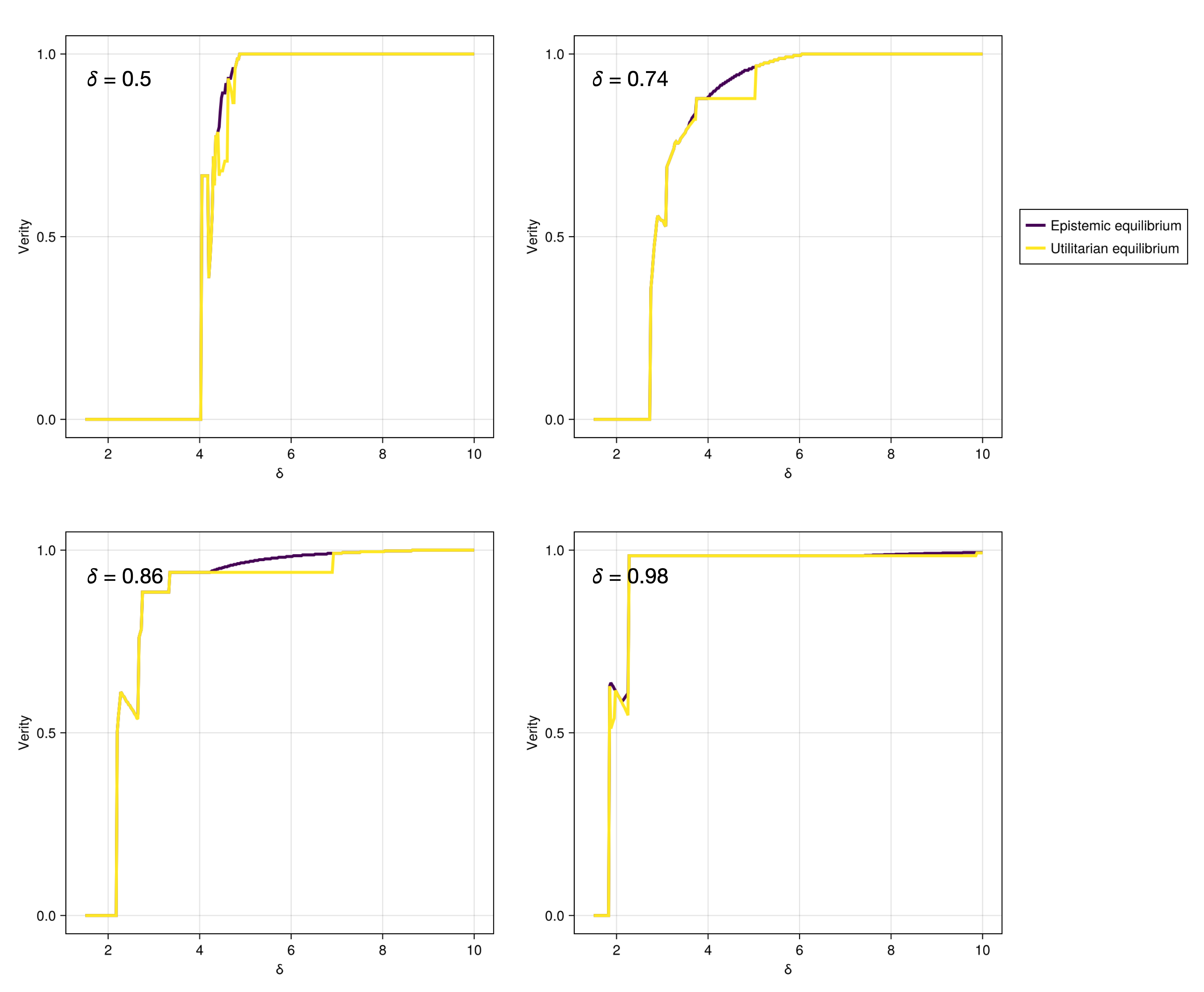}
\caption{Collective accuracy, for $n = 3$ and various delta, plotted
as a function of debaters' truth-weights. We compare the result of using
the epistemic equilibrium selection criterion versus a more familiar
utilitarian criterion. Aside from the broad similarity of the results,
this also verifies that the non-monotonicity of accuracy as a function
of truth-weight is not an artefact of our equilibrium selection
criterion.}
\end{figure}

\hypertarget{supplementary-references}{%
\section*{Supplementary References}\label{supplementary-references}}
\addcontentsline{toc}{section}{Supplementary References}

\hypertarget{refs}{}
\begin{CSLReferences}{0}{0}
\leavevmode\vadjust pre{\hypertarget{ref-betz2013debate}{}}%
\CSLLeftMargin{1. }
\CSLRightInline{G. Betz, \emph{Debate dynamics: {How} controversy
improves our beliefs} (Springer Netherlands, 2013).}

\end{CSLReferences}

\end{document}